\documentclass{aa}  

\usepackage{array}
\usepackage{graphicx}
\usepackage{multirow}
\usepackage{color}
\usepackage{hyperref}
\usepackage{makecell}
\usepackage[capitalize]{cleveref}
\usepackage{txfonts}
\renewcommand{\arraystretch}{1.2}

\begin{document}

   \title{First detection of ultra-fast outflows in a quiescent galaxy}

   \authorrunning{Y. Xu et al.}
   
   \author{Yerong Xu\inst{1,2}\corrauth{yerong.xu@ice.csic.es},
            Malgorzata Siudek\inst{3}  \email{malgorzata.siudek@gmail.com},
            Victor Rodríguez Morales\inst{1} \email{victorproyecto98@gmail.com},
            Mar Mezcua \inst{1,2} \email{mezcua@ice.csic.es},
            Hai-cheng Feng\inst{4} \email{hcfeng@ynao.ac.cn},
            Ciro Pinto\inst{5} \email{ciro.pinto@inaf.it},
            James N. Reeves\inst{6,7} \email{james.n.reeves456@gmail.com},
            Stefano Bianchi\inst{8} \email{stefano.bianchi@uniroma3.it},
            Valentina Braito\inst{6,7,9} \email{valentina.braito@inaf.it},
            \and  Luigi C. Gallo\inst{10} \email{luigi.gallo@smu.ca}
        }
        
   \institute{Institute of Space Sciences (ICE, CSIC), Campus UAB, Carrer de Magrans, 08193 Barcelona, Spain
   \and Institut d'Estudis Espacials de Catalunya (IEEC), Edifici RDIT, Campus UPC, 08860 Castelldefels, (Barcelona), Spain
   \and Instituto de Astrof\'isica de Canarias (IAC), Departamento de Astrof\'isica, Universidad de La Laguna (ULL), 38200, La Laguna, Tenerife, Spain
   \and Yunnan Observatories, Chinese Academy of Sciences, Kunming 650216, Yunnan, People’s Republic of China
   \and INAF - IASF Palermo, Via U. La Malfa 153, I-90146 Palermo, Italy
    \and Institute for Astrophysics and Computational Sciences, Department of Physics, The Catholic University of America, Washington, DC 20064, USA
   \and INAF, Osservatorio Astronomico di Brera, Via Bianchi 46, I-23807 Merate (LC), Italy
   \and Dipartimento di Matematica e Fisica, Università degli Studi Roma Tre, via della Vasca Navale 84, I-00146 Roma, Italy
   \and Dipartimento di Fisica, Universit\`a di Trento, Via Sommarive 14, Trento 38123, Italy
    \and Department of Astronomy \& Physics, Saint Mary's University, 923 Robie Street, Halifax, NS B3H 3C3, Canada
             }

   \date{Received XXX; accepted XXX}


  \abstract
   {Outflows in active galactic nuclei (AGN) are recognized as a fundamental mechanism driving the co-evolution of supermassive black holes (SMBHs) and their host galaxies by regulating the gas reservoir for the SMBH growth and the host star formation. Although powerful outflows are frequently detected in gas-rich, active star-forming galaxies, their existence and potential impact within gas-poor, quiescent galaxies remain poorly understood. }
   {We report the first detection of a powerful ultra-fast outflow (UFO) in a nearby quiescent galaxy KUG 1208+386, providing a multiscale analysis of AGN winds from nuclear to galactic scales.}
   {
   We performed a Bayesian-based X-ray spectroscopy of archival \textit{XMM-Newton} and \textit{NuSTAR} observations to characterize the circumnuclear materials and UFOs. Using optical spectra from the Dark Energy Spectroscopic Instrument (\textit{DESI}), we constrained the kinematics of the galactic-scale outflows and the star formation history (SFH) of the host galaxy. Additionally, we derived the galaxy properties, such as stellar mass ($M_\star$) and star formation rate (SFR), via multi-wavelength spectral energy distribution (SED) fitting of photometric data.
   }
   {We detect a nuclear X-ray UFO with a velocity of $v_{\rm out} \approx -0.07c$ and a kinetic power of $\dot{E}_{\rm UFO}=(0.8\mbox{--}6.5) \times 10^{43}$\,erg/s. This power is sufficient to drive effective AGN feedback ($\dot{E}_{\rm UFO}/L_\mathrm{Edd}=(1\mbox{--}8)\%$) and far exceeds the galactic [OIII] outflow power $\sim 10^{40}$\,erg/s. Host galaxy analysis reveals a massive quiescent system (specific star formation rate $\sim3\times10^{-12}\,\mathrm{yr}^{-1}$) that has been quenched $\sim$9\,Gyr ago. The central AGN is obscured by a line-of-sight (LOS) column density of $\log (N_\mathrm{H}^\mathrm{LOS}/\mathrm{cm}^{-2})
   \sim23$ and the circumnuclear scattering material is Compton-thick $\log( N_\mathrm{H}^\mathrm{scatter}/\mathrm{cm}^{-2})=24.7^{+0.8}_{-0.5}$, indicating a dense nuclear environment.
   }
   {
   The discovery of a nuclear UFO in a long-quenched massive galaxy challenges the paradigm that UFOs are exclusive to gas-rich, star-forming systems, suggesting instead that they are governed by the local circumnuclear environments, rather than the global gas reservoir. Our results indicate that episodic, powerful winds can maintain the quiescent state of KUG 1208+386 over Gyr timescales, supporting a wind-driven `maintenance' mode of AGN feedback that is distinct from the classical jet mode. Future deep X-ray and sub-mm observations are essential to definitively characterize the cold gas reservoir and the coupling efficiency among multi-phase and multi-scale outflows.
   }


   \maketitle
   \nolinenumbers
%

\section{Introduction}\label{sec:intro}

Active Galactic Nuclei (AGN) are widely considered to play a pivotal role in shaping the evolution of their host galaxies. The observed correlations between the mass of supermassive black holes (SMBHs) and host galaxy properties, such as the $M_\mathrm{BH}\mbox{--}\sigma_\star$  relation ($\sigma_\star$ is the stellar dispersion velocity), suggest a co-evolutionary process regulated by AGN feedback \citep[e.g.,][]{2013Kormendy,2018Harrison,2024Harrison}. This feedback typically manifests through outflows that can either expel/heat or compress the interstellar medium (ISM), thus suppressing or enhancing the star formation of the host galaxy \citep[e.g.,][]{2012Fabian,2014Zubovas,2017Maiolino}.

Among various manifestations of feedback, ultra-fast outflows (UFOs) are the most extreme AGN winds. Characterized by mildly relativistic velocities ($v\gtrsim 10000$\,km/s) and typically detected via strongly blueshifted absorption lines in X-ray spectra \citep[e.g.,][]{2010Tombesi}, UFOs are thought to be launched from the innermost regions of the accretion disk. These winds are likely accelerated by radiation pressure \citep{2000Proga,2016Nomura,2017Matzeu,2021Xu,2023Xu} or magnetic fields \citep{2010Fukumura,2015Fukumura,2025Gu,2026Reeves}, with recent high-resolution insights from missions like \textit{XRISM} suggesting a potential hybrid mechanism even in high-accretion AGN \citep{2025Xrism,2026Mizumoto,2026Reevesa}. Crucially,
UFOs can carry sufficient kinetic power, often exceeding the threshold ($\gtrsim 0.5\%$--$5\% L_\mathrm{Edd}$, $L_\mathrm{Edd}$ is the Eddington luminosity) required to exert global influence on their host galaxies \citep{2005DiMatteo,2010Hopkins}. 

Such powerful winds have been predominantly observed in AGN \citep[$30\mbox{--}40\%$ detection rate;][]{2010Tombesi,2013Gofford,2021Chartas,2023Matzeu}, classified as Seyfert galaxies or quasars depending on the AGN contribution to the galaxy luminosity. Most host galaxies of these AGN are gas-rich galaxies with non-negligible star formation activity, i.e., star-forming (SF) or green-valley (GV) galaxies, characterized by the specific star formation rate $\mathrm{sSFR}\equiv\mathrm{SFR}/M_\star\geq10^{-11}\,\mathrm{yr}^{-1}$, where SFR is the star formation rate and $M_\star$ is the galaxy stellar mass \citep[e.g.,][]{2013Veilleux,2015Tombesi,Koss2021,2024Travascio}.  The abundant gas reservoirs can fuel the AGN accretion ($\lambda_\mathrm{Edd}\equiv L_\mathrm{bol}/L_\mathrm{Edd}\gtrsim0.01$, $L_\mathrm{Bol}$: bolometric luminosity) necessary to launch winds.

However, the presence and role of powerful AGN winds in quiescent galaxies remain poorly understood. Most massive galaxies in the local Universe reside on the `red sequence', characterized by old stellar populations and negligible star formation \citep[$\mathrm{sSFR<10^{-11}\,\mathrm{yr}^{-1}}$;][]{2007Faber}. The recent discovery of an unexpected abundance of massive quiescent systems at $z>3$ by \textit{JWST} \citep[e.g.,][]{2023Carnall,2023CarnallNat} further emphasizes the need for a continuous heating source to prevent halo gas from cooling and re-igniting star formation over Gyr timescales. Current models typically invoke `radio-mode' feedback, where mechanical energy from collimated jets in low-luminosity AGN \citep[LLAGNs, $\lambda_\mathrm{Edd}<0.01$;][]{Yuan2014} balances cooling in the ISM \citep{2012Fabian}. Nonetheless, it remains an open question whether wide-angle winds can provide an alternative mechanism to jets for sustaining quiescence and regulating the $M_\mathrm{BH}\mbox{--}\sigma_\star$ relation long after galaxy formation.

Besides nuclear and short-lived X-ray UFOs, broad (i.e., exceeding the galaxy dynamics) [OIII] $\lambda5008\AA$ emission has become one of the most widely used diagnostics of galactic-scale outflows, associated with Myr-timescale AGN activity \citep[e.g.,][]{2014Harrison}. As a forbidden transition, the [OIII] $\lambda5008\AA$ emission line is suppressed in the high-density sub-pc scales of the AGN broad-line region (BLR) and thus serves as a clean tracer of gas on pc to kpc scales in the narrow-line region (NLR) \citep[e.g.,][]{2005Boroson,2005Greeneb,2013Mullaney}. While the bulk of the outflowing mass often resides in the molecular phase, [OIII] offers unparalleled advantages in tracking the kinematics and energetics of kpc-scale outflows due to its high luminosity in optical spectra \citep[e.g.,][]{2018Fischer}.

Nevertheless, establishing a direct observational link between nuclear winds and galactic outflows remains a challenge. The primary limitation is the detection of nuclear winds \citep[e.g.,][]{2013Veilleux}, requiring high signal-to-noise (SNR) ultraviolet (UV) / X-ray spectroscopy from space-based observatories. The UFO detection rate and its transient nature \citep[e.g.,][]{2023Reeves,2025Gu} further complicate the detection. Consequently, by far only a small number of systems ($\sim10$) have been confirmed to host UFOs and kpc-scale winds simultaneously. Studies of these rare sources have revealed complex outflow behaviors, from momentum-conserving \citep[IRAS F11119+3257, PG 1211+143, I Zw 1;][]{2015Tombesi,2013Pounds,2019Reeves} to energy-conserving winds \citep[Mrk 231, IRAS 17020+4544][]{2015Feruglio,2018Longinotti}, with several exceptions deviating from both modes \citep[APM 08279+5255, PDS 456, MCG-03-58-007;][]{2017Feruglio,2018Braito,2019Sirressi,2019Bischetti,2024Travascio,2025Xrism}, highlighting the need for more case studies.

KUG 1208+386 (hereafter KUG 1208) is a nearby \citep[$z=0.022974$, i.e. $d\sim100$\,Mpc;][]{2022Koss}, radio-quiet \citep[$R_\mathrm{1.4GHz}=5.9\,\mathrm{mJy}$,][]{2016Wong}, Seyfert 1.2 galaxy. It hosts a SMBH with a mass of $\mathrm{log}(M_\mathrm{BH}/M_\odot)=7.5\pm0.5$ accreting at a bolometric luminosity of $\log(L_\mathrm{bol}/\mathrm{erg\,s}^{-1})\simeq44.22$ with an Eddington ratio of $\lambda_\mathrm{Edd}\simeq0.035$ \citep{2022Koss}. It is one of the targets in \textit{Swift}/BAT All-sky Hard X-ray Survey and has a bright hard X-ray flux of $F_\mathrm{14-195\,keV}\simeq2.2\times10^{-11}\,\mathrm{erg/s/cm^{-2}}$ with corresponding luminosity of $\log(L_\mathrm{14-195\,keV}/\mathrm{erg\,s}^{-1})\simeq43.42$ \citep{2025Lien}. In this work, we discover a powerful nuclear UFO and a galactic-scale [OIII] wind in KUG 1208+386, and identify that it has remained quiescent for $\sim$9\,Gyr. This source provides a rare laboratory for investigating the role of
AGN winds in the evolution of quiescent galaxies.

Section \ref{sec:data-reduction} describes the reduction of data utilized in this work. In Section \ref{sec:spectral-modeling} and \ref{sec:MW}, we show our spectral analysis of X-ray and other multi-wavelength data, respectively. The physical properties of outflows are derived in Section \ref{sec:ionized-winds}. We discuss the implications in Section \ref{sec:discussion} and show conclusions in Section \ref{sec:conclusions}. In this paper, we assume the cosmological constants at $H_\mathrm{0}=70$\,km/s/Mpc, $\Omega_\mathrm{\Lambda}=0.73$ and $\Omega_\mathrm{M}=0.27$. 

\section{Data Preparation}\label{sec:data-reduction}

KUG 1208 was observed by \textit{XMM-Newton} for 15.8ks in 2009 (ObsID: 0601780901), by \textit{NuSTAR} for 31.7ks in 2019 (ObsID: 60061225002), and by \textit{Swift} for 27.2ks in 2007 (ObsID: 00037139001-4) and 5.6ks in 2019 (ObsID: 00080074001). It was also monitored by \textit{DESI} twice in 2022 for 860s (TargetID: 2305843039240720934) and 1130s (TargetID: 39633044046350357). Multi-wavelength photometric surveys further cover KUG 1208 from the far-ultraviolet (FUV) to the mid-infrared (MIR).

\subsection{\textit{XMM-Newton}}\label{subsec:xmm-reduction}
\textit{XMM-Newton} data were reduced following standard threads with the \textit{XMM-Newton} Science Analysis System (SAS v20.0.0) and calibration files available by September 2025. We reduced the EPIC-MOS1/2 and EPIC-pn data using \textsc{emproc} and \textsc{epproc}, respectively. Time intervals affected by solar flares were excluded by applying separate thresholds for EPIC-MOS and EPIC-pn, namely $<0.8$ and $<5$ counts/s in the $10\mbox{--}12$\,keV band. The abnormally high threshold for EPIC-pn is because of the bright background contamination when EPIC was operated in the Full-Frame mode. The net exposure times are 7.8ks and 7.3/8.0ks for EPIC-pn and EPIC-MOS1/2, respectively. The source spectra were extracted from a circular region with a radius of 30 arcsec centered on the source, while the background spectra were extracted from a region with a radius of 60 arcsec offset but close to the source. No severe pile-up effect was found. The $2\mbox{--}10$\,keV count rates of pn, MOS1, and MOS2 are 0.24, 0.07, and 0.08 cts/s, respectively. RGS data were not utilized due to the background domination.
The Optical Monitor (OM) photometry data were also reduced with \textsc{omichain} to constrain the UV/optical band of the spectral energy distribution (SED). KUG 1208 was observed in the UVW1 (2910\AA) and UVM2 (2310\AA) filters during observations. The response files are retrieved from the ESA webpage\footnote{https://www.cosmos.esa.int/web/xmm-newton/om-response-files}. 

\subsection{\textit{NuSTAR}}\label{subsec:NuSTAR-reduction}
The reduction of the \textit{NuSTAR} data was conducted following the standard procedures with the NuSTAR Data Analysis Software (NUSTARDAS v.2.1.4) and the updated calibration files (CALDB v20250224). The source spectrum was extracted from a circular region of radius 60 arcsec, and the background spectrum from a free-source region with a radius of 120 arcsec. The $3\mbox{--}40$\,keV count rates for both FPMA and FPMB are 0.14 cts/s.

\subsection{\textit{Swift}}\label{subsec:Swift-reduction}
Only \textit{Swift}/UVOT data were employed in this paper to help constrain the SED of AGN in KUG 1208. Four filters: UVW2 (1928\AA), UVM2 (2246\AA), UVW1 (2600\AA), and U (3465\AA), were utilized in 2007 observations, while the 2019 observation used six filters, with additional B (4392\AA) and V (5468\AA) filters. The UV/optical photometry data were processed following the standard threads with the \textsc{uvotsource} tool. The source and background extraction regions are circles with a radius of 5 arcsec and 20 arcsec, respectively. The photometry data were transferred into the XSEPC-readable format using \texttt{uvot2pha} with the canned response files\footnote{https://heasarc.gsfc.nasa.gov/docs/heasarc/caldb/data/swift/uvota/\\ index.html}. No significant flux variations ($<10\%$) were found among \textit{Swift}/UVOT 2007 and 2019 exposures, and \textit{XMM-Newton}/OM 2009 observations within the same filters. Therefore, we took their average fluxes for analysis.

\subsection{\textit{DESI}}\label{subsec:DESI-reduction}

\begin{figure}[htbp]
    \centering        
    \includegraphics[width=\columnwidth,height=7cm]{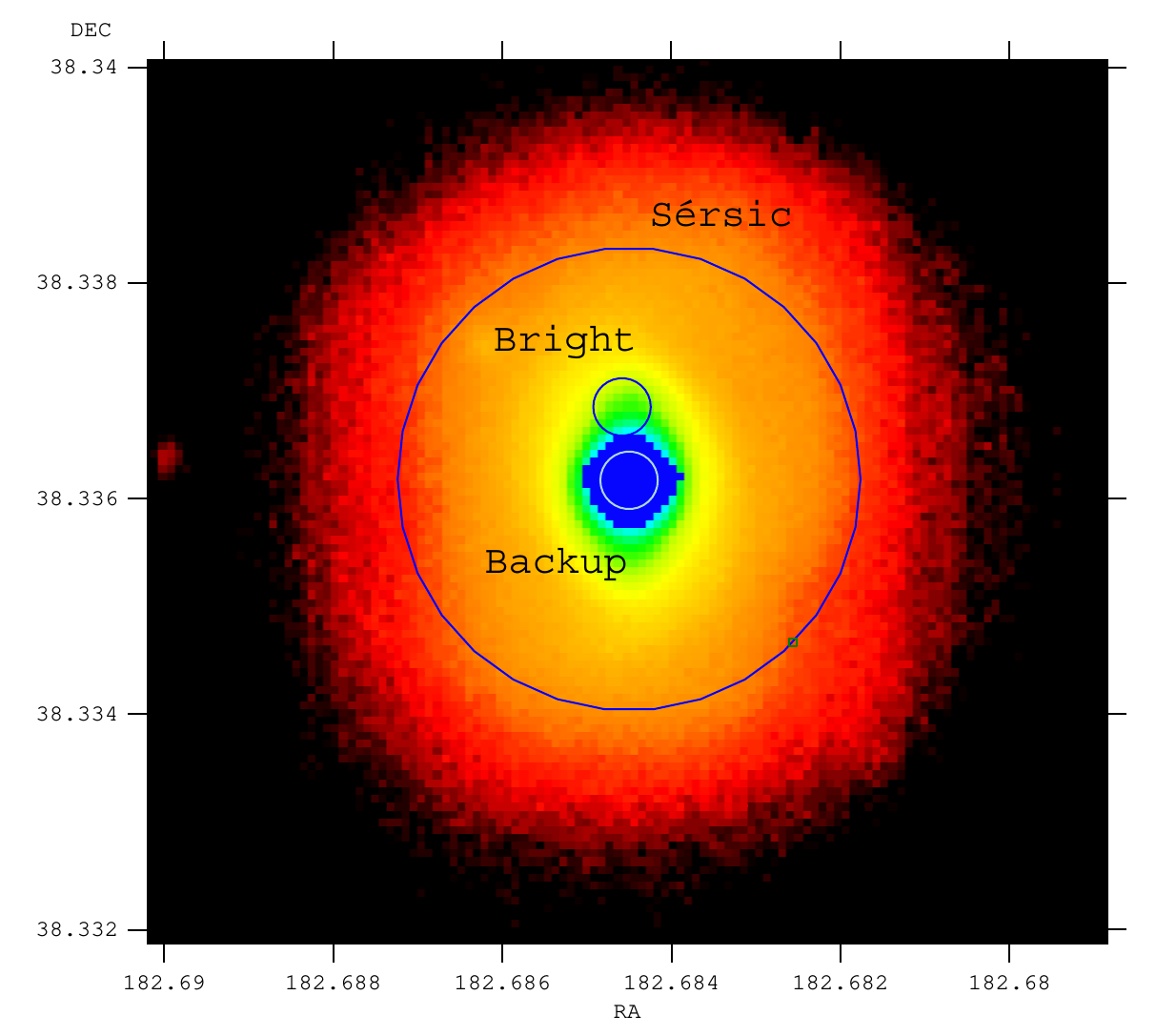}
    \caption{The image from the Dark Energy Camera Legacy Survey \citep[DECaLS;][]{2019Dey} DR9 of KUG 1208+386. The FWHM/2 of the PSF (radius of 0.96" in the $g$ band) of \textit{DESI} spectra in the `bright' and `backup' catalogs are marked by circles, where the `Backup' spectrum points to the AGN and `Bright' spectrum is offset. The S\'ersic region (a radius of 7.73") is also shown to illustrate the extent of the host galaxy.
    }
    \label{fig:DECaLS}
\end{figure}

We retrieved \textit{DESI} spectra from Data Release 1 \citep[DR1,][]{2024DESI}. KUG 1208 was observed twice in April 2022 in `bright' (TargetID: 39633044046350357) and `backup' (TargetID: 2305843039240720934) modes separately, with a temporal separation of less than one month. The difference in the overall flux normalization (by a factor of $\sim$5) and emission lines (AGN lines in the `backup' spectrum) between the two spectra is attributed to variations in the pointing (see Fig.\ref{fig:DECaLS}), where the `Backup' spectrum points to the AGN and the `Bright' spectrum is offset, pointing to the galaxy of KUG 1208. The half of the full width at half maximum (FWHM) of the point spread function (PSF) in the $g$-band of each pointing is indicated in Fig.\ref{fig:DECaLS} (where $\mathrm{FWHM}(\mathrm{PSF_\mathrm{g}})/2=0.96$"\footnote{https://www.legacysurvey.org/dr9/files/}), along with the S\'ersic radius (the half-light radius of the galaxy, $R_\mathrm{S\acute{e}rsic}=7.73$").

\subsection{UV-Optical-MIR photometry}\label{subsec:photometry}
Archival multi-wavelength photometry data were retrieved for the measurement of galaxy properties: far- and near-UV (\textit{FUV} and \textit{NUV}) measurements from the Galaxy Evolution Explorer \citep[GALEX DR5;][]{2011Bianchi}; Sloan Digital Sky Survey (SDSS DR16) magnitudes in the optical/IR
bands \citep[\textit{u, g, r, i, z};][]{2020Jonsson}; 
IR photometry in the \textit{J, H}, and \textit{K} bands from the Extended Source Catalog of The Two Micron All Sky Survey \citep[2MASS/XSC;][]{2003Cutri,2006Skrutskie}; and MIR photometry (\textit{W1$\mbox{--}$W4}) from the Wide-field Infrared Survey Explorer \citep[WISE;][]{2010Wright,2021Cutri}. 
Galactic dust extinction was corrected using the relation $A_\lambda=R_\lambda\times\mathrm{E(B-V)}$, adopting $\mathrm{E(B-V)}=0.0173$ \citep{2011Schlafly}. The extinction coefficients $R_\lambda$ were assigned based on the wavelength regime: values for the GALEX, SDSS, and 2MASS+WISE bands were taken from \citet{2019Wall}, \citet{2011Schlafly}, and \citet{2019Wang}, respectively.

\section{X-ray Spectral Analysis}\label{sec:spectral-modeling}

\textit{XMM-Newton} and \textit{NuSTAR} spectra (5 in total) were jointly analyzed using \textsc{PyXSPEC} \citep[v2.1.4,][]{1996Arnaud,2021Gordon}. We considered the EPIC-pn spectrum between $2\mbox{--}10$\,keV, EPIC-MOS spectra between $2\mbox{--}9$\,keV, and FPM spectra between $3\mbox{--}40$\,keV in our analysis to avoid the background contamination (see raw data against the continuum model in Fig.\ref{app:fig:raw_data}). All spectra were optimally binned with \textsc{ftgrouppha} \citep{2016Kaastra} and fitted with C-statistic \citep{1979Cash}. Galactic absorption was explained by \texttt{tbabs} with a column density fixed at $N_H^\mathrm{MW}=1.65\times10^{20}\,\mathrm{cm}^{-2}$  \citep{2016HI4PI} using the solar abundances calculated by \citet{2000Wilms}. The host galaxy absorption was accounted for by \texttt{phabs}. The calibration and flux differences were considered by a variable \texttt{constant}. The redshift was fixed at $z=0.022974$ \citep{2022Koss} with \texttt{zashift}, and X-ray luminosities were derived via \texttt{clumin}. 

The significance of additional components was examined using the Akaike Information Criterion corrected for the small sample size \citep[$\mathrm{AIC_c}$;][]{1974Akaike,2016Emmanoulopoulos}, where $\mathrm{AIC_c}=2p+\mathrm{C\mbox{--}stat}+2p(p+1)/(n-p-1)$, $n$ is the number of data bins and $p$ is the number of free parameters. The $\mathrm{AIC_c}$ difference between two models, for example, $\Delta\mathrm{AIC_c}=\mathrm{AIC_c^A}-\mathrm{AIC_c^B}$, implies model A is $e^\mathrm{-\Delta\mathrm{AIC_c}/2}$ times likely than model B, where $\Delta\mathrm{AIC_c}<-10$ (i.e. $>99.3\%$ significance) represent a decisive model preference. The model parameter space was explored with \textsf{nautilus} \citep{2023Lange}, a Bayesian importance nested sampler, using 1000 live points and 4 neural networks. Unless specifically stated, all parameter uncertainties reported in this work correspond to the 68\% highest-density intervals of the posterior distributions. For each free parameter, we adopted either linear or log-linear uniform priors and ran the sampler until 1\% of the evidence remained in the live set, with an effective sample size of 10,000. Bayesian analysis also provides an alternative way to compare models with the Bayesian evidence (logZ), where model A is considered decisively better than model B if $\Delta\mathrm{logZ}=\mathrm{logZ}_{A}-\mathrm{logZ}_{B}>2$, following the criteria of \citet{1961jeffreys} and \citet{1995kass}.

\begin{figure}[htbp]
    \centering        
    \includegraphics[width=\columnwidth]{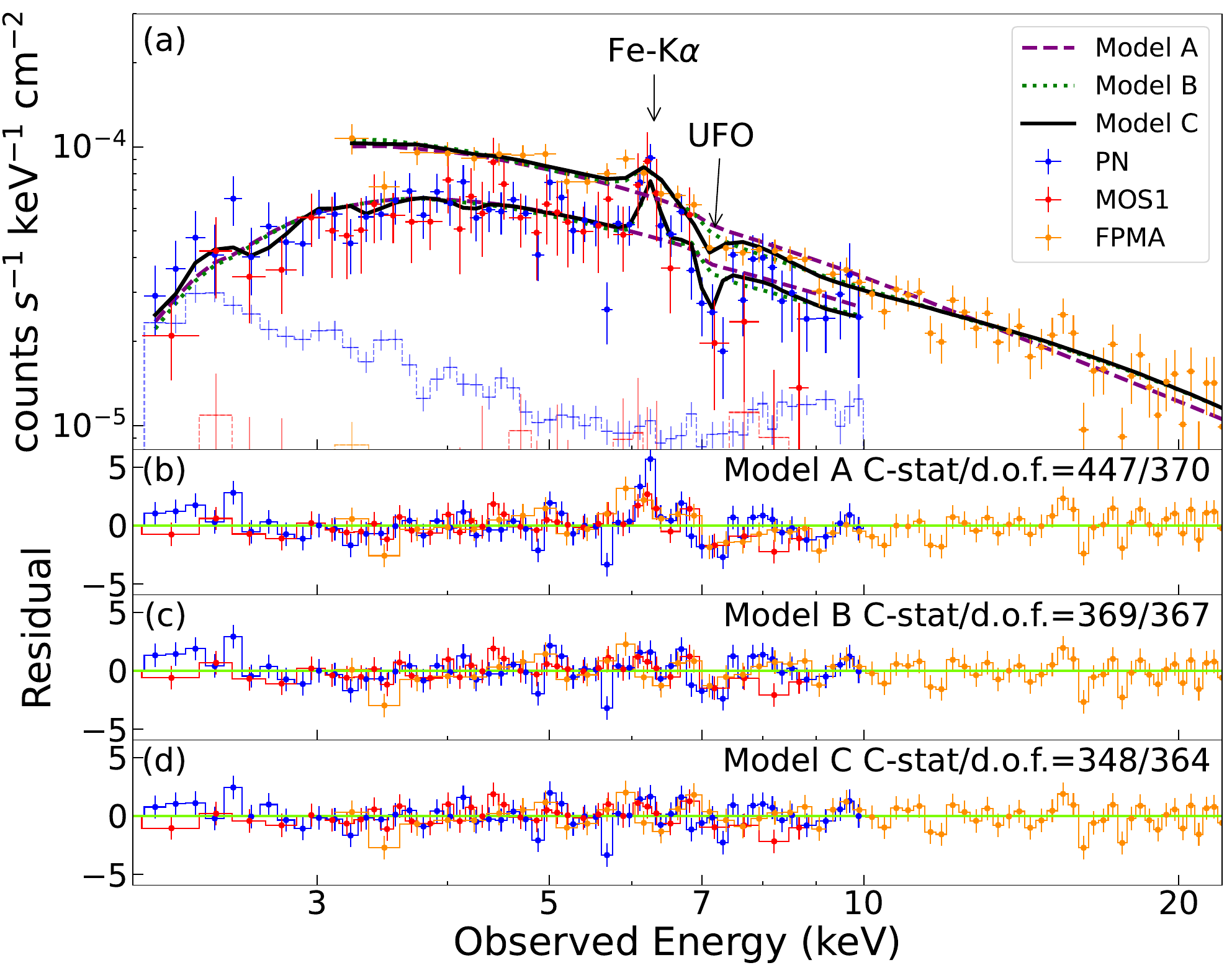}
    \caption{Source X-ray spectra (\textit{a}) and residuals (\textit{b-d}) with respect to model A-C. For clarity, only data from pn (\textit{blue}), MOS1 (\textit{red}), and FPMA (\textit{orange}) are shown, together with their corresponding background spectra (faint, diluted histograms in matching colors). The best-fit models of Model A-C are shown in \textit{dashed purple, dotted green}, and \textit{solid black} lines, respectively, with relevant fit statistics in residual panels. The Fe-K$\alpha$ emission and the UFO absorption line are annotated. 
    }
    \vspace{-2mm}
    \label{fig:spectrum+chi}
\end{figure}

We tried various model combinations to explain the spectra. Their best-fit models (with residuals) and parameters are summarized in Fig.\ref{fig:spectrum+chi} and Tab.\ref{tab:fits}, respectively.

\begin{table*}
\setlength{\tabcolsep}{3pt}
\renewcommand{\arraystretch}{1.3}
\centering
\caption{Best-fit parameters of different model combinations with respect to \textit{XMM-Newton} joint \textit{NuSTAR} spectra. 
}
\begin{tabular}{lcccccc}
\hline
\hline
Description & Model & Parameter &Unit & Model A$^a$ & Model B$^b$ & Model C$^c$ \\
\hline
Galactic Absorption & {\tt tbabs} &    $N^\mathrm{Gal}_\mathrm{H}$ & $10^{20}$ cm$^{-2}$      & \multicolumn{3}{c}{$1.65^{\star}$} \\
\hline
Redshift & {\tt zashift} &    $z$ &      & \multicolumn{3}{c}{$0.022974^{\star}$} \\
\hline
Host absorption & {\tt phabs} &    $N^\mathrm{host}_\mathrm{H}$ & $10^{22}$ cm$^{-2}$      & $10.4^{+0.5}_{-0.6}$ & $11.0^{+0.9}_{-0.8}$ & $9.9^{+0.8}_{-0.8}$ \\
\hline
&  &    $\Gamma_\mathrm{XMM}$ &       & $1.43^{+0.03}_{-0.03}$ & $1.62^{+0.08}_{-0.07}$ & $1.51^{+0.08}_{-0.07}$ \\
&  &    $\Gamma_\mathrm{Nu}$ &       & $1.63^{+0.02}_{-0.02}$ & $1.89^{+0.07}_{-0.07}$ & $1.81^{+0.06}_{-0.04}$ \\
Hot Corona & {\tt nthComp} &    kT$_{e}$ &    keV   &  & $100^{\star}$ &  \\
&  & kT$_\mathrm{bb}$ & keV & &  $0.1^{\star}$ & \\
& &    log(Norm$_{HC}$) &      & $-3.14^{+0.03}_{-0.03}$ & $-3.02^{+0.07}_{-0.07}$ & $-2.96^{+0.05}_{-0.04}$ \\
\hline
 & &  log($N_\mathrm{H}^\mathrm{Tor}$) &    $\mathrm{cm}^{-2}$   &  & $24.4^{+0.8}_{-0.2}$ & $24.7^{+0.8}_{-0.5}$   \\
 & & $A_\mathrm{Fe}$ & $Z_\odot$ & & \multicolumn{2}{c}{$1^\star$} \\
Torus & \texttt{borus12}  &    $\cos\theta$ & & & \multicolumn{2}{c}{$0.5^\star$}  \\
&  &    $C_F$ &      &  & $0.54^{+0.11}_{-0.08}$ & $0.46^{+0.14}_{-0.10}$   \\
& &    log(Norm$_{Tor}$) &      & & $-2.64^{+0.16}_{-0.15}$ & $-2.71^{+0.14}_{-0.13}$ \\
\hline
\multirow{6}{1em}{Wind} & \multirow{6}{4em}{\texttt{PION\_ABS}} &    log$\xi^\mathrm{wind}$ &  erg cm s$^{-1}$    &  &  & $3.3^{+0.2(0.4)\dagger}_{-0.2(0.5)\dagger}$  \\
 &  &    $N_\mathrm{H}^\mathrm{wind}$ &  $10^{24}\,\mathrm{cm}^{-2}$    &  &  & $0.35^{+0.13(0.27)\dagger}_{-0.12(0.23)\dagger}$  \\
  &  &    $N_\mathrm{H;cor}^\mathrm{wind\,\ddagger}$ &  $10^{24}\,\mathrm{cm}^{-2}$    &  &  & $0.41^{+0.15(0.31)\dagger}_{-0.14(0.27)\dagger}$  \\
 &  &    $\sigma_{v}^\mathrm{wind}$ &  km/s    &  &  & $1000^{\star}$  \\
  &  &    $z^\mathrm{wind}_\mathrm{LOS}$ &  $c$ (3$\times10^{5}$ km/s)    &  &  & $-0.070^{+0.006(0.022)\dagger}_{-0.008(0.026)\dagger}$  \\
  &  &    $v^\mathrm{wind\,\ddagger}_\mathrm{LOS}$ &  $c$ (3$\times10^{5}$ km/s)    &  &  & $-0.072^{+0.006(0.023)\dagger}_{-0.009(0.028)\dagger}$  \\
\hline
\multirow{2}{1em}{Luminosity} & \multirow{2}{4em}{\texttt{clumin}} &  log$L_\mathrm{2\mbox{--}10keV}^\mathrm{EPIC}$  &  \multirow{2}{1em}{erg/s}    & & $42.92^{+0.08}_{-0.10}$   \\
 &  &  log$L_\mathrm{2\mbox{--}10keV}^\mathrm{FPM}$  &      &  & $43.08^{+0.11}_{-0.15}$  \\
\hline
\multirow{2}{4em}{Cross-calibration}  & \multirow{2}{4em}{\texttt{constant}}  &  $C_\mathrm{PN}$/$C_\mathrm{MOS1}$/$C_\mathrm{MOS2}$   &   &   & $1^{\star}/0.93^{+0.09}_{-0.09}/1.06^{+0.09}_{-0.09}$  \\
&  &  $C_\mathrm{FPMA}$/$C_\mathrm{FPMB}$  &   &   & $1.53^{+0.16}_{-0.23}/1.53^{+0.17}_{-0.23}$  \\
\hline
Fit-statistics &  &  C-stat/d.o.f.  &   & 447/370 &369/367  & 348/364   \\
\multicolumn{2}{l}{Corrected Akaike Information Criterion}  & $\mathrm{AIC}_{c}$  &   & 463 &391  & 380   \\
Bayesian Evidence  & & logZ  &   & -241.5 & -209.0   & -205.0  \\
\hline
\hline
\end{tabular}
\label{tab:fits}
\begin{flushleft}
{$^a$ Model A: \texttt{constant*tbabs*zashift*phabs*clumin*nthComp}}\\
{$^b$ Model B: \texttt{constant*tbabs*zashift*phabs*clumin*(nthComp+borus12)}}\\
{$^c$ Model C: \texttt{constant*tbabs*zashift*phabs*PION\_ABS*clumin*(nthComp+borus12)}}\\
{$^{\star}$ The parameter is fixed.}\\
{$^{\dagger}$ Values in parentheses correspond to the 2$\sigma$ uncertainty.}\\
{$^{\ddagger}$ The parameter is corrected by the special relativity following \citet{2020Luminari,2021Luminari}.}\\
\end{flushleft}
\vspace{-4mm}
\end{table*}

\subsection{\textit{Model A}}\label{subsec:model-A}

The X-ray spectrum was first explained by an absorbed Comptonization (power-law-like) model (\texttt{nthComp}), i.e., \texttt{constant*tbabs*zashift*phabs*clumin*nthComp}, referred to as Model A. To account for decade-long spectral evolution, the photon index $\Gamma$ was allowed to vary between \textit{XMM} and \textit{NuSTAR} data. Given the lack of constraints, we fixed the temperature of seed photons (kT$_\mathrm{bb}=0.1$\,keV) and hot corona (kT$_\mathrm{e}=100$\,keV) \citep[e.g.][]{2017Fabian}. The soft X-ray excess component, commonly observed in Type 1 AGN \citep[e.g.,][]{2004Gierli,2009Bianchi,2021XuSE,2024Waddell}, was not included in our analysis due to the exclusion of $<2$\,keV data, highly affected by the background.

The host galactic absorption column density along our line-of-sight (LOS) was obtained at $N_\mathrm{H}^\mathrm{host}\sim10^{23}\,\mathrm{cm}^{-2}$, compatible with results in \citet{2013Vasudevan}, making it an obscured AGN falling in the Compton-thin regime \citep[$N_\mathrm{N}^\mathrm{LOS}=10^{22}\mbox{--}1.5\times10^{24}\,\mathrm{cm}^{-2}$; e.g.][]{1999Risaliti,2015Brandt,2017Ricci}.

A strong Fe-K$\alpha$ emission line ($\sim6.4$\,keV) and a tentative absorption feature ($\sim7$\,keV) appear in the residual spectra of both EPIC and NuSTAR data (panel $b$ of Fig.\ref{fig:spectrum+chi}). We applied two line search methods to estimate the significance of these lines with results shown in Fig.\ref{app:fig:line} (see details in Appendix.\ref{app:sec:raw_data}). The Fe K$\alpha$ line is significantly detected at $>3.9\sigma$ in both methods, with its significance peaking for a narrow core ($\sigma\leq1000$\,km/s). An absorption feature is located at $7.25^{+0.07}_{-0.07}$\,keV in the rest frame with a significance of $\sim5\sigma$ and $\sim3\sigma$ in single-trial and true significance (including the look-elsewhere effect) spectra, respectively. Its energy excludes the possibility of the Fe-K edge from ISM absorption \citep[$7.11\mbox{--}7.13$\,keV,][]{2018Rogantini}, which implies a UFO with a velocity of $v>0.04c$ assuming its origin is from Fe XXV/XXVI.

\subsection{\textit{Model B}}\label{subsec:model-B}

The Fe-K emission line implies the reprocessing of X-ray photons emitted from the hot corona off colder materials \citep[e.g.,][]{1989Fabian,2005Ross}. Due to its narrow line width, $\sigma<0.14$\,keV (derived from fitting a Gaussian line), it is likely that the Fe-K emission line originates from the distant reflection from the SMBH, rather than from the inner accretion disk \citep[e.g.,][]{1995Mushotzky,2009Fabian}. Therefore, we included a physical torus model \texttt{borus12} \citep{2018Balokovic,2019Balokovic} into Model A for the X-ray reflection, named as Model B. The \texttt{borus12} model calculates the transmitted and scattered X-ray photons in the cold, neutral, and uniformly distributed gas with a toroidal geometry. The photon index and the electron gas temperature were linked to those of the hot Comptonization. The angle of inclination of the torus is fixed at $\theta=60^\circ$ (i.e., cos$\theta=0.5$) due to its poor constraint. The scattering column density ($N_\mathrm{H}^\mathrm{Tor}$) and the covering factor ($C_\mathrm{F}$) of the torus were left free. 

The inclusion of a torus model significantly improves the fitting statistics by $\Delta\mathrm{C-stat}=78$ with 3 degrees of freedom (d.o.f.). Both $\Delta\mathrm{AIC}_{c}=-72$ and $\Delta\mathrm{logZ}=32.5$ strongly favor the presence of a distant reprocessing component. The scattering column density, log$(N_\mathrm{H}^\mathrm{Tor}/\mathrm{cm}^{-2})=24.4^{+0.8}_{-0.2}$, indicates an intrinsic Compton-thick torus. The covering factor is $C_\mathrm{F}=0.54^{+0.11}_{-0.08}$, broadly consistent with typical values found for AGN tori \citep[e.g.,][]{2019Marchesi,2021Zhao}.
However, we caution that the posterior distribution (see Fig.\ref{app:fig:corner} for the reference) shows that the covering fraction $C_\mathrm{F}$ is not constrained at the $3\sigma$ level, while $N_\mathrm{H}^\mathrm{Tor}$ remains in the Compton-thick regime over the same confidence range. Therefore, while the detailed torus geometry cannot be determined with the present data, the fit supports the presence of dense distant reprocessing material.

The alternative distant reflection model from the accretion disk \citep[\texttt{xillverCp};][]{2010Garc,2013Garc} was also tested, yielding a statistically indistinguishable fit. We, therefore, keep the torus interpretation.

\subsection{\textit{Model C}}\label{subsec:model-C}
To physically explain the Fe absorption line, we employed the advanced photoionization code \texttt{PION} in the SPEX package \citep[v3.08.01,][]{1996Kaastra,2015Miller,2016Mehdipour}. This model self-consistently calculates the transmission and emission spectra of photoionized gas by a given radiation field, where the photoionization equilibrium is sensitive to the shape of irradiating SEDs \citep[e.g.,][]{2020Pinto,2025Xu,2025Buhariwalla}.

To constrain the intrinsic SED, we incorporated the UV and optical photometry alongside a disk blackbody component into Model B, accounting for both Galactic and host-galaxy dust extinction (see details in Appendix.\ref{app:sec:SED}). The constrained SED yielded an ionizing luminosity ($1\mbox{--}1000$\,Ryd) of $L_\mathrm{ion}\sim 2.4\times10^{43}\,\mathrm{erg/s}$ and a bolometric luminosity ($1\,\mathrm{eV}\mbox{--}1000\,\mathrm{keV}$) of $L_\mathrm{bol}\sim1.5\times10^{44}\,\mathrm{erg/s}$, which aligns with the value inferred from \textit{Swift}/BAT data \citep{2022Koss}. Furthermore, we verified that a potential unmodeled soft X-ray excess only marginally increases the wind ionization parameter ($\Delta\log\xi \sim 0.15$) and does not alter our main conclusions. We therefore adopted the fitted SED (Fig.\ref{app:fig:SED}) for all subsequent analyses.

The SED was then input into \texttt{PION} to calculate synthetic transmission spectra. To implement the \texttt{PION} model in XSPEC, we adopted the code used in \citet{2019Parker} to construct the tabulated XSPEC version of \texttt{PION}. Here, we only included the absorption component, i.e., assuming the opening angle at $\Omega=0$ and a full LOS covering $C_\mathrm{F}=1$, called \texttt{PION\_ABS}, and the model combination is named as model C. \texttt{PION\_ABS} is characterized by four parameters: column density $N_\mathrm{H}^\mathrm{wind}$, ionization parameter log$\xi^\mathrm{wind}$, turbulence velocity $\sigma_{v}^\mathrm{wind}$, and the LOS shift of gas $z^\mathrm{wind}_\mathrm{LOS}$. Due to limited data quality and spectral resolution, the turbulent velocity was fixed at $\sigma_v^\mathrm{wind}=1000$\,km/s (Fig.\ref{app:fig:line}). Photoionization parameters were linked across \textit{XMM} and \textit{NuSTAR} spectra as they remain consistent within uncertainties.

The best-fit spectra, posteriors, and parameters are shown in Fig.\ref{fig:spectrum+chi}, \ref{app:fig:corner}, and Tab.\ref{tab:fits}, respectively. The addition of this wind component is remarkably supported by statistics, $\Delta\mathrm{C-stat}=-21$, $\Delta\mathrm{AIC_c}=-11$, and $\Delta\mathrm{logZ}=4.0$. The posterior distribution shows a partial degeneracy between the photoionization parameters, especially $N_\mathrm{H}^\mathrm{wind}$, and the LOS shift $z_\mathrm{LOS}^\mathrm{wind}$, due to the limited CCD spectral resolution. Therefore, the origin of the dominant UFO absorption feature, either Fe XXV or Fe XXVI, cannot be uniquely determined, although the best-fit solution slightly favors Fe XXV. To reflect this uncertainty, we report UFO parameters with both the 68\% uncertainties and the 95\% uncertainties in parentheses throughout the following analysis. The best-fit model requires an ionization parameter $\log(\xi^\mathrm{wind}/\mathrm{erg\,cm\,s}^{-1})=3.3^{+0.2(0.4)}_{-0.2(0.5)}$, a column density $N_\mathrm{H}^\mathrm{wind}=3.5^{+1.3(2.7)}_{-1.2(2.3)}\times10^{23}\,\mathrm{cm}^{-2}$, and a LOS redshift $z^\mathrm{wind}_\mathrm{LOS}=-0.070^{+0.006(0.022)}_{-0.008(0.026)}$. After applying special relativistic corrections \citep{2020Luminari,2021Luminari}, we derive a LOS velocity of $v^\mathrm{wind}_\mathrm{LOS}=-21600^{+1800(6900)}_{-2700(8400)}\,\mathrm{km/s}$ and a column density of $N_\mathrm{H;cor}^\mathrm{int}=4.1^{+1.5(3.1)}_{-1.4(2.7)}\times10^{23}\,\mathrm{cm}^{-2}$, marking the first UFO identification in KUG 1208.


\section{UV-optical-MIR Data Analysis}\label{sec:MW}
In this section, we present a comprehensive analysis of optical \textit{DESI} spectra and multi-wavelength photometric data to characterize galactic-scale outflows and host properties of KUG 1208.

\subsection{Optical Line Fitting}\label{subsec:optical-line-spectra}
\textit{DESI} spectra ($3600\mbox{--}7500\AA$) were modeled using \textsf{PyQSOFit} \citep{2018Guo, 2019Shen}. The code performs the fit with $\chi^2$ statistics in the rest frame after automatically correcting for the Galactic extinction using the \citet{1998Schlegel} dust map. The spectrum is modeled as a superposition of an AGN power-law, Fe II emission, various Gaussian emission lines, and a host stellar component fitted with simple stellar population (SSP) templates from the E-MILES library \citep{2016Vazdekis}. Parameter uncertainties are estimated via Markov chain Monte Carlo (MCMC) sampling.

For the line fittings, the velocity shifts and dispersions of all narrow lines (e.g., [OIII] $\lambda\lambda5008, 4960\AA$, [NII] $\lambda\lambda$6585, 6550$\AA$, and [SII]$\lambda\lambda$ 6718, 6733$\AA$, and the narrow cores of Balmer lines) were tied, assuming they originate from NLR with similar kinematics. Coronal lines (e.g. [FeVII] $\lambda$3760, $\lambda$5721, $\lambda$6087$\AA$) were also linked with the narrow lines to ensure the fitting stability given their limited SNR. For the Balmer series, broad components of H$\alpha$ and H$\beta$ were modeled using three sets of Gaussians, with their dispersions and shifts tied between two lines for physical consistency. Additionally, an outflow component with free kinematics was included for the [OIII] doublet to probe galactic-scale outflows. Following theoretical quantum ratios, the flux ratios of [OIII] $\lambda5008/\lambda4960\AA$ and [NIII] $\lambda6585/\lambda6550\AA$ were fixed at 2.95:1, and 3.06:1, respectively.


The PyQSOFit decomposition results for the AGN and offset-AGN spectra are presented in Fig.\ref{fig:DESI-backup} and \ref{app:fig:DESI-bright}, respectively, along with zoom-in plots of the H$\alpha$ and H$\beta$ regions. The peak wavelengths, velocity dispersions, and Galactic-extinction-corrected fluxes for the lines of interest are summarized in Tab.\ref{app:tab:lines_complex}. Generally, moving from the on-AGN to the offset-AGN pointing, both the line fluxes and the AGN continuum contribution decrease, and coronal lines are no longer detectable. In both spectra, the H$\alpha$ line is well modeled by one narrow, two intermediate, and one broad component, which are classified based on their velocity dispersions ($\sigma<300$\,km/s, $300-2000$\,km/s and $>2000$\,km/s, respectively). Although Balmer lines can sometimes trace galactic outflows \citep[e.g.,][]{2024Travascio}, the non-narrow Balmer components in KUG 1208 may not serve this purpose. Their dispersions remain nearly constant across both pointings (a separation of $\sim1.2$\,kpc), indicating that these broad and intermediate components originate from the PSF spillover of BLR emission, rather than a kpc-scale outflow.

Including an [OIII] outflow component significantly improves the fit for both spectra ($\Delta\chi^2/\mathrm{d.o.f.}=-3075/3$ and $-229/3$ for the AGN and offset-AGN spectra, respectively), confirming the presence of [OIII] outflows \citep[e.g.][]{2013Liu}. The velocity dispersion of this component decreases from $\sigma_\mathrm{[OIII];O}=240\pm4$\,km/s at the core to $\sigma_\mathrm{[OIII];O}=187\pm11$\,km/s at the offset location, supporting a physical origin for the kpc-scale [OIII] outflows, rather than an instrumental artifact.

\subsubsection{Host Extinction}\label{subsubsec:host-extinction}
We utilized the narrow-line Balmer decrement to estimate the dust extinction on the gas in the host galaxy. Assuming the intrinsic H$\alpha$/H$\beta$ ratio of 2.86, corresponding to Case B recombination for the NLR \citep[characterized by a temperature of $T=10^4$K and an electron density of $n_\mathrm{e}=10^2\,\mathrm{cm}^{-3}$;][]{2006Osterbrock}, we derived the color excess of $\mathrm{E(B-V)_\mathrm{gas}}=0.23\pm0.02$ using the \citep{1989Cardelli} extinction law with $R_V=3.1$. A consistent value of $\mathrm{E(B-V)_\mathrm{gas}} = 0.24 \pm 0.12$ was also obtained from the offset-AGN spectrum, albeit with larger uncertainties. However, we caution that using the Balmer decrement as a proxy for gas extinction entails intrinsic systematic uncertainties, particularly given the potential for complex dynamical or geometric variations of the BLR \citep[e.g., see Fig.1 in][]{2024Li}.

\subsubsection{Black Hole Mass Estimation}\label{subsubsec:BH-mass}

We evaluated the black hole mass utilizing PyQSOFit results of the \textit{DESI} AGN spectrum (Tab.~\ref{app:tab:lines_complex}), taking advantage of its SNR of the broad Balmer profiles. We employed the virial mass scaling relations from \citet{2005Greene}. For the H$\alpha$ line, the mass is derived as:
\begin{equation}
\begin{split}
    M_\mathrm{BH}\mathrm{(H\alpha;b)}&=(2.0^{+0.4}_{-0.3}\times10^{6})\left(\frac{L_\mathrm{H\alpha;b}}{10^{42}\mathrm{erg/s}}\right)^{0.55\pm0.02} \\
    & \left(\frac{\mathrm{FWHM}_\mathrm{H\alpha;b}} {10^3\mathrm{km/s}}\right)^{2.06\pm0.06}\,M_\odot, 
\end{split}
\end{equation}
while the H$\beta$-derived mass is given by:
\begin{equation}
\begin{split}
M_\mathrm{BH}\mathrm{(H\beta;b)}&= (3.6^{+0.2}_{-0.2}\times10^{6})\left(\frac{L_\mathrm{H\beta;b}}{10^{42}\mathrm{erg/s}}\right)^{0.56\pm0.02} \\
 &\left(\frac{\mathrm{FWHM}_\mathrm{H\beta;b}}{10^3\mathrm{km/s}}\right)^2\,M_\odot, 
\end{split}
\end{equation} 
where $L_\mathrm{H\alpha(\beta);b}$ and $\mathrm{FWHM}_\mathrm{H\alpha(\beta);b}$ are the luminosity and FWHM of the broad component of H$\alpha(\beta)$, respectively.
Given that the broad Balmer lines require three Gaussian components, we assumed that all broad components originate from the BLR regardless of their complex kinematics. Consequently, the line luminosity is the sum of these components. To account for the multi-component profile, where a conventional FWHM measurement only captured the intermediate component and missed the dispersion of the broad base, we utilized the effective line dispersion, defined as $\mathrm{FWHM}=2\sqrt{2\ln2}\sigma_\mathrm{eff}\sim2.355\sqrt{\sum F_i\sigma_i^2/\sum F_i}$, where $\sigma_\mathrm{eff}$ is the flux weighted dispersion \citep{2004Peterson}, and $(F_i, \sigma_i)$ are line flux and dispersion of non-narrow components, respectively. The resulting values are $\mathrm{FWHM}_\mathrm{H\alpha}\sim5700\,$km/s and $\mathrm{FWHM}_\mathrm{H\beta}\sim6200\,$km/s. These measurements yield consistent results from H$\alpha$ ($\log M_\mathrm{BH}/M_\odot\sim7.8$) and H$\beta$ ($\log M_\mathrm{BH}/M_\odot\sim7.7$) with uncertainties dominated by intrinsic systematics of the single-epoch virial approach ($\sim0.5$\,dex). Thus, the masses derived from both Balmer lines are consistent with the result of \citet{2022Koss}. For subsequent analysis, we adopted the H$\alpha$-derived value, $\log(M_\mathrm{BH}/M_\odot)=7.8\pm0.5$, due to its superior SNR. Combining with the bolometric luminosity derived in Sec.~\ref{subsec:model-C} ($L_\mathrm{bol}\sim1.5\times10^{44}\mathrm{erg/s}$), we calculated an Eddington ratio of $\lambda_\mathrm{Edd}=0.019^{+0.040}_{-0.013}$, placing KUG 1208 in the typical Seyfert regime, characterized by geometrically thin accretion disks.

\subsection{SED Fitting with CIGALE}\label{subsec:SED}
We measured the galaxy properties (e.g. stellar mass and SFR) through the Code Investigating GALaxy Emission \citep[CIGALE v2025.1;][]{2019Boquien,2020Yang,2022Yang}, which is designed to use a library of models to fit the SED of galaxies considering the energy balance conservation. Our input data comprised the FUV-optical-MIR photometries detailed in Sec.\ref{subsec:photometry} and the hard X-ray flux ($2\mbox{--}10$\,keV) from the X-ray spectroscopy. We adopted a delayed star formation history (SFH) and SSP models from \citet{2003Bruzual} assuming an initial mass function from \citet{2003Chabrier}. We incorporated the standard nebular emission model of \citet{2011Inoue} and the dust attenuation model based on the modified \citet{2000Calzetti} starburst attenuation curve. The reprocessed dust emission was characterized by the \citet{2014Draine} model. We assumed solar metallicity through the fitting. The AGN contribution was explained by the \citet{2006Fritz} models, while the X-ray emission was self-consistently integrated using the module developed by \citet{2020Yang}, which connects the X-ray luminosity to the AGN UV/optical emission.

The best-fit broadband SED and the corresponding residuals are shown on the left panel of Fig.\ref{app:fig:host}. CIGALE yields a satisfactory fit with a reduced $\chi^2$ of 3.3. The primary residuals manifest as an excess in the WISE2 (4.6$\mu$m) band, a discrepancy likely stemming from the complex MIR emission characteristic of passive galaxies. This excess can be attributed to contamination from circumstellar dust associated with evolved stars (e.g., asymptotic giant branch stars). The galaxy stellar mass and SFR are well-constrained at $M_\star = (3.1 \pm 0.6) \times 10^{10}\,M_\odot$ and $\mathrm{SFR} = 0.09 \pm 0.05\,M_\odot\,\mathrm{yr}^{-1}$, respectively. While the absence of FIR/sub-mm data may lead to an overestimation of SFRs, typically by a factor of two \citep{2022Kim}, this potential bias does not alter our conclusions. These parameters imply an exceptionally low specific SFR of $\mathrm{sSFR}\equiv \mathrm{SFR}/M_\star \approx 3 \times 10^{-12}\,\mathrm{yr}^{-1}$, categorizing KUG 1208 as a massive, quenched galaxy residing on the red sequence \citep[e.g.,][]{2006Weinmann,2007Schiminovich,2007Salim}.

Furthermore, the inferred stellar attenuation, $\mathrm{E(B-V)_\mathrm{star}}\sim0.3$, is comparable to the gas extinction derived from the Balmer decrement. This similarity suggests a diffuse and relatively uniform dust distribution across the galaxy, a hallmark of quiescent galaxies \citep[e.g.,][]{2011Wild,2014Price}. In such systems, dense clouds associated with active star formation have largely dissipated, leaving both stars and ionized gas to be attenuated by the same diffuse ISM.

\subsection{Stellar Population Analysis}\label{subsec:ppxf}

The stellar kinematics and populations of KUG 1208 were obtained based on the AGN-subtracted offset spectrum from the \textsf{PyQSOFit} decomposition, since it has less contamination from AGN and coronal lines. Following the procedures in \citet{2024Spiniello}, we performed spectral fitting using the Penalized PiXel-Fitting code \citep[pPXF v9.4.2;][]{2004Cappellari,2017Cappellari,2023Cappellari}. pPXF decomposes an observed spectrum into a combination of SSP templates of different ages and metallicities, convolved with an LOS velocity distribution. The resulting weights represent the light contribution from stars formed across different cosmic epochs, allowing for the SFH reconstruction. 

We performed the fit over a wavelength range of $3500\mbox{--}7500\AA$. The SSP models from the E-MILES stellar library \citep{2016Vazdekis} were convolved to match the \textit{DESI} resolution (right panel of Fig.\ref{app:fig:host}). We first used pPXF to determine the stellar kinematics. In this step, we adopted additive Legendre polynomials with DEGREE $=14$, where DEGREE controls the order of the additive continuum correction. This value was adopted because the stellar velocity dispersion $\sigma_\star$ becomes insensitive to further increases in polynomial degree. The fit yields a negligible stellar velocity ($V_\star <10$\,km/s) and a velocity dispersion of $\sigma_\star = 92\pm5$\,km/s, where the uncertainty was estimated via a bootstrap procedure, in which the polynomial degree was varied over DEGREE $=6\mbox{--}23$.

\begin{figure}[]
    \centering        
    \includegraphics[width=\columnwidth]{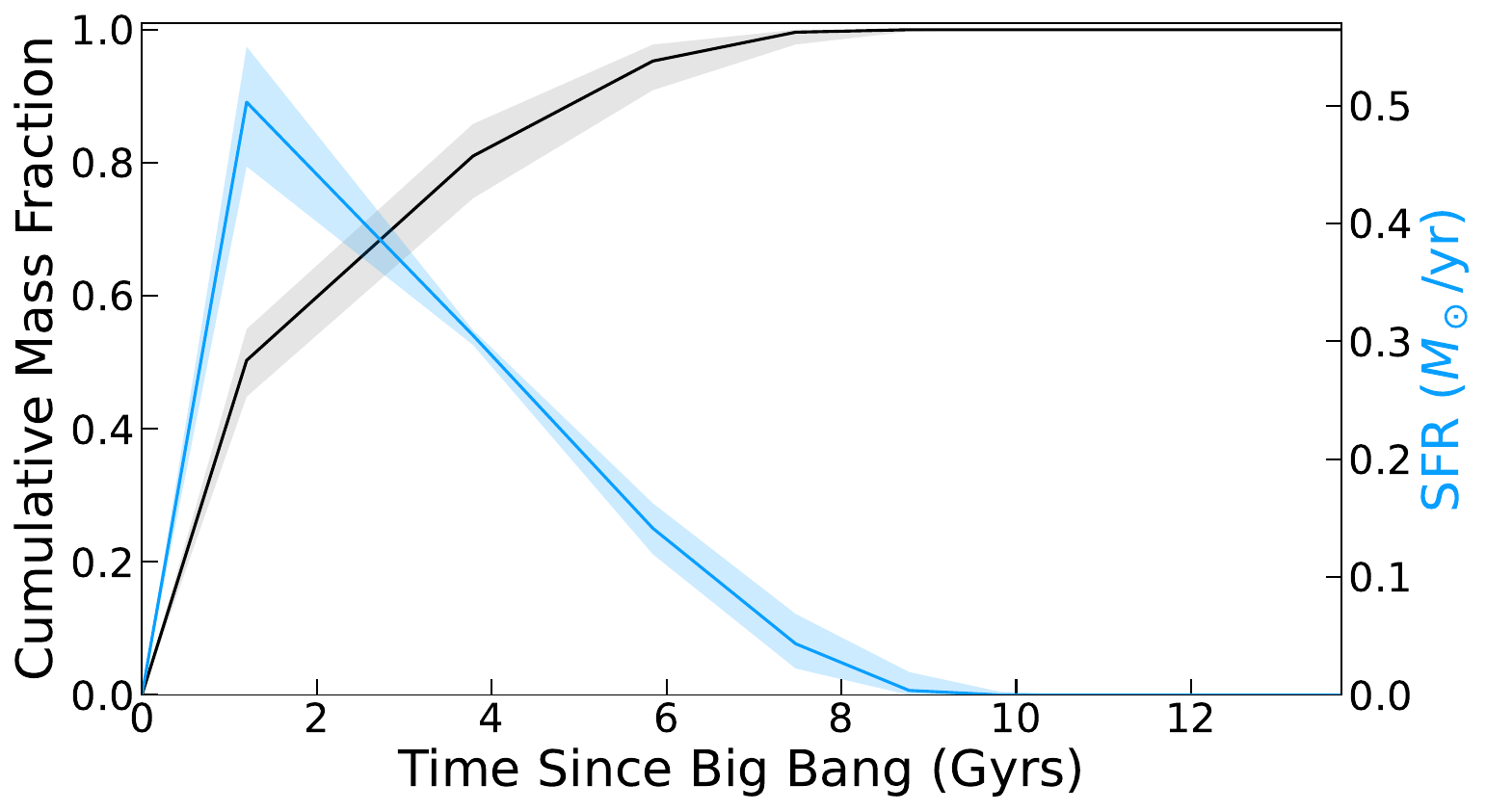}
    \caption{
    History of the regularized cumulative mass (\textit{left}) and the star formation (\textit{right}). 90\% of the stellar mass was assembled within $4.6\pm0.7$\,Gyr post-BB. SFR peaks at $\sim1.2$\,Gyr post-BB. The \textit{solid} line shows the 50th percentile, and \textit{shaded} regions represent the $1\sigma$ uncertainty.  
    }
    \vspace{-4mm}
    \label{fig:ppxf-fit}
\end{figure}

Following the standard pPXF procedure for stellar population analysis\footnote{https://github.com/micappe/ppxf\_examples/tree/main}, we then re-ran pPXF to infer the stellar population properties and star-formation history. The stellar kinematics were fixed to the best-fitting values obtained above. In this second run, additive polynomials were replaced with multiplicative Legendre polynomials with MDEGREE $=10$ to preserve the relative absorption line strengths. The MDEGREE value was chosen because stable results were obtained for $8<\mathrm{MDEGREE}<20$ \citep{2021Spiniello}. We first performed multiple unregularized fits (REGUL$=$0) to determine the maximum regularization parameter (MAX\_REGUL$=7.2$) by rescaling the noise until $\chi^2$ increased by $\Delta\chi^2=\sqrt{\rm 2\times d.o.f.}$. Then we included the regularization to obtain a physically smooth SFH. Parameter uncertainties were estimated using a bootstrap routine (256 iterations), where random noise was added to the spectrum while the continuum treatment (MDEGREE $=8\mbox{--}20$) and regularization strength (between MAX\_REGUL-5 and MAX\_REGUL) were also perturbed. These uncertainties therefore include statistical noise and part of the setup dependence associated with the pPXF hyperparameters. The cumulative mass fraction as a function of time since the Big Bang (BB) and SFH are shown in Fig.\ref{fig:ppxf-fit}. We found that KUG 1208 assembled 90\% of its stellar mass within $4.6\pm0.7$\,Gyr, with a peak SFR at $\sim$1.2\,Gyr post-BB, implying that most of the stellar mass was formed early. We note that the detailed SFH (the peak and amplitude of the SFR) may depend mildly on the adopted SSP library and residual AGN subtraction, while the main conclusion remains unchanged that KUG 1208 is dominated by an old stellar population, consistent with its quiescent state inferred from the SED fitting in Sec.\ref{subsec:SED}.

\section{Physical Properties Of Ionized Winds}\label{sec:ionized-winds}

\subsection{UFOs}\label{subsec:UFOs}

Following the procedure in the literature \citep[e.g.,][]{2013Tombesi,2015Gofford,2025Xiang}, we constrained the UFO location based on two assumptions: 1) the radial thickness of the wind does not exceed its distance from the central source, $\Delta R^\mathrm{wind}\leq R^\mathrm{wind}$; and 2) the observed outflow velocity exceeds the local escape velocity of the accretion disk, $v^\mathrm{wind}\geq v^\mathrm{esc}$. Given the definition of the ionization parameter ($\xi\equiv L_\mathrm{ion}/n_\mathrm{H}R^2$) and the relation $N_\mathrm{H}^\mathrm{wind}=n_\mathrm{H}^\mathrm{wind}\Delta R^\mathrm{wind}\leq f_\mathrm{V}n_\mathrm{H}^\mathrm{wind} R^\mathrm{wind}$, the first assumption yields an upper limit of the wind radius:
\begin{equation}\label{eq:high-limit-R_UFO}
R^\mathrm{wind}_\mathrm{max}=\frac{L_\mathrm{ion}f_\mathrm{V}}{\xi^\mathrm{wind} N^\mathrm{wind}_\mathrm{H}}=r_1 f_\mathrm{V},
\end{equation}
where $f_\mathrm{V}$ is the volume filling factor ($f_\mathrm{V}=1$ for a homogeneous outflow) and $r_1=L_\mathrm{ion}/\xi N_\mathrm{H}$. Conversely, the second assumption provides a lower limit of the UFO location:
\begin{equation}\label{eq:min-R_UFO}
    R^\mathrm{wind}_\mathrm{min}=\frac{2GM_\mathrm{BH}}{(v^\mathrm{wind})^2}=\frac{2GM_\mathrm{BH}}{(v^\mathrm{wind}_\mathrm{LOS})^2}\mathrm{cos}^2\theta=r_2\mathrm{cos}^2\theta,
\end{equation}
where $G$ is the gravitational constant, $\theta$ is the inclination angle of the outflow stream relative to the LOS, and $r_2=2GM_\mathrm{BH}/v_\mathrm{LOS}^2$. For the same absorber that satisfies both assumptions, the physical consistency requires that the lower limit of the wind radius must not exceed its upper limit, $R_\mathrm{min}^\mathrm{wind}\leq R_\mathrm{max}^\mathrm{wind}$. For a small inclination angle ($\mathrm{cos}\theta\sim1$), the lower limit on the volume filling factor can be $f_\mathrm{V}^\mathrm{min}=r_2/r_1$. For non-zero inclination angles, $R_\mathrm{min}^\mathrm{wind}$ would be further reduced, yet this method offers a valuable approach for constraining UFO locations.

Using the spectroscopic results in Sec.\ref{subsec:model-C} and Tab.\ref{tab:fits}, we inferred $r_1 = 1.0^{+0.9(3.0)}_{-0.4(0.7)}\times10^{-2}$\,pc ($\approx3300\,R_\mathrm{g}$) and $r_2 = 1.2^{+0.2(1.4)}_{-0.2(0.6)}\times10^{-3}$\,pc ($\approx383\,R_\mathrm{g}$), respectively, where $R_\mathrm{g}=GM_\mathrm{BH}/c^2$ is the gravitational radius. All asymmetric uncertainties were determined via Monte Carlo (MC) sampling of a split-normal distribution with $10^6$ iterations. The resulting lower limit for the filling factor is  $f_\mathrm{V}^\mathrm{min}=0.11^{+0.09(0.33)}_{-0.05(0.08)}$, comparable with the clumpy UFOs in PDS 456 \citep[$f_\mathrm{V}=0.1\mbox{--}0.3$;][]{2025Xrism}.

The mass outflow rate is estimated as 
\begin{equation}\label{eq:mass-outflow-rate-UFO}
    \dot{M}_\mathrm{out}=4\pi C_\mathrm{F}\mu m_\mathrm{p}\frac{L_\mathrm{ion}}{\xi}v^\mathrm{wind}f_\mathrm{V}f_\mathrm{cov} ,
\end{equation}
where $f_\mathrm{cov}\equiv\Omega/4\pi$ is the opening fraction assumed at 0.4 \citep[from the UFO detection fraction in AGN; e.g.][]{2013Tombesi}, $\mu=1.23$ is the mean atomic weight and $m_\mathrm{p}$ is the proton mass. Assuming a unity volume filling factor ($f_\mathrm{V}=1$), the mass outflow rate is $\dot{M}_\mathrm{out}^\mathrm{max} = 4^{+3(10)}_{-2(3)}\,M_\odot\,\mathrm{yr}^{-1}$, generally exceeding the Eddington mass accretion rate $\dot{M}_\mathrm{Edd}\equiv L_\mathrm{Edd}/\eta c^2\sim1.4\,M_\odot/\mathrm{yr}$ \citep[assuming a radiative accretion efficiency of $\eta=0.1$;][]{1973Shakura}. Even adopting the minimal volume filling factor, the inferred rate $\dot{M}_\mathrm{out}^\mathrm{min} = 0.5^{+0.2(0.5)}_{-0.2(0.3)}\,M_\odot\,\mathrm{yr}^{-1}$ remains comparable to $\dot{M}_\mathrm{Edd}$. 
Our conservative estimate demonstrates that the UFO carries a significant fraction of the available accreting matter, potentially providing substantial kinetic feedback on the host galaxy.

The kinetic power of a wind, assuming it reaches a terminal velocity without further significant acceleration, is:
\begin{equation}\label{eq:kinetic-energy-UFO}
    \dot{E}_\mathrm{K}=0.5\dot{M}_\mathrm{out}(v^\mathrm{wind})^2.
\end{equation}
For the unity and minimal filling factor cases, the UFO kinetic power are $\dot{E}_\mathrm{K}^\mathrm{max}=6^{+5(22)}_{-3(5)}\times10^{44}\,\mathrm{erg/s}$ and $\dot{E}_\mathrm{K}^\mathrm{min}=8^{+3(7)}_{-3(5)}\times10^{43}\,\mathrm{erg/s}$, respectively, corresponding to $\dot{E}_\mathrm{K}^\mathrm{max}/L_\mathrm{Edd}=8^{+6(27)}_{-3(6)}\%$, and $\dot{E}_\mathrm{K}^\mathrm{min}/L_\mathrm{Edd}=1.0^{+0.4(1.0)}_{-0.3(0.6)}\%$, both generally exceeding the classical 0.5\% benchmark for efficient AGN feedback \citep[e.g.,][]{2005DiMatteo,2010Hopkins}. Since this benchmark was mainly motivated by quasar-mode feedback in star-forming systems, we also compared our results with simulations for quiescent galaxies. Simulations of feedback in quiescent galaxies usually focus on whether the AGN input energy (via jets) could balance the halo gas cooling \citep[e.g.,][]{Ciotti2017,Eisenreich2017}. Using the empirical relation from \citet{Best2006}, $\bar{H}=10^{28.4}(M_\mathrm{BH}/M_\odot)^{1.6}$\,erg/s, the time-averaged heating rate required to balance the cooling for $M_\mathrm{BH}\sim10^{7.8}\,M_\odot$ is $\sim7.6\times10^{40}\,\mathrm{erg/s}$, far below the UFO kinetic power, suggesting UFOs could, in principle, sustain the quiescent state, though the transfer efficiency and AGN duty cycle remain uncertain.

To evaluate the driving mechanism, we compared the outflow momentum rate $\dot{P}_\mathrm{out}=\dot{M}_\mathrm{out}v^\mathrm{wind}$ with the radiation momentum rate $\dot{P}_\mathrm{rad}=L_\mathrm{bol}/c$. The inferred radiative momentum rate is $\dot{P}_\mathrm{rad}\sim5\times10^{33}\,\mathrm{dyn}$, while the UFO momentum flux ranges from $\dot{P}_\mathrm{out}^\mathrm{min}=7^{+3(5)}_{-2(5)}\times10^{34}\,\mathrm{dyn}$ (for $f_\mathrm{V}^\mathrm{min}$) to $\dot{P}_\mathrm{out}^\mathrm{max}=6^{+4(16)}_{-2(4)}\times10^{35}\,\mathrm{dyn}$ (for $f_\mathrm{V}=1$), depending on the choice of filling factors. The $1\mbox{--}2$ order-of-magnitude momentum boost indicates that radiation pressure alone is insufficient to drive the outflow, suggesting a magnetically accelerated UFO, as expected for a sub-Eddington system.

\subsection{[OIII] $\lambda5008\AA$ Outflows}\label{subsec:OIII-winds}

We characterized the warm ionized outflow using the [OIII]$\lambda5008$ line. The outflow velocity was characterized by $W_{80}\equiv v_{90}-v_{10}$, the velocity width containing 80\% of the emission line flux, where $v_{90}$ and $v_{10}$ are velocities at the 10th and 90th percentiles of the total flux, respectively \citep[e.g.,][]{2013Liu,2014Harrison,2019Manzano-King,2025Rodriguez}. The outflow velocity was defined as $v_\mathrm{out}=W_{80}/1.3$ \citep[suitable for the spherically symmetric or wide-angle bi-cone outflow models; outlined in][]{2013Liu}. For AGN-centered and offset-AGN spectra, we obtained, respectively obtained $W_{80}=410\pm6$\,km/s and $W_{80}=375\pm20$\,km/s, resulting in $v_\mathrm{out}=315\pm4$\,km/s and $v_\mathrm{out}=288\pm15$\,km/s.

The [OIII] electron density $n_\mathrm{e}$ was estimated from the density-sensitive flux ratio of the collisionally excited [SII]  $\lambda6718/\lambda6733\AA$ doublet. Here, we implicitly assumed that the ionized gas [SII] and [OIII] are generally co-spatial and share a common density. Using the \texttt{PyNeb} package \citep{2013Luridiana} to solve the gas equilibrium equations with an assumption of a typical nebular electron temperature of $T_e = 10^4$\,K, we obtained $n_\mathrm{e}=625\pm60\,\mathrm{cm}^{-3}$ and $n_\mathrm{e}=103\pm90\,\mathrm{cm}^{-3}$, respectively from centered and offset spectra, where uncertainties were calculated through standard error propagation.

The spatial extent of the [OIII] outflow was constrained by its detection in the offset-AGN spectrum, located at a projected distance of $\sim2.5$" from the nucleus (corresponding to $R_\mathrm{out}^\mathrm{min}\sim1.15$kpc). This detection provides a lower limit on the outflow radius, assuming an isotropic geometry. As a conservative upper bound, we assumed the ionized outflow to remain within the main stellar body of the host galaxy and adopted the Sérsic radius $R_\mathrm{out}^\mathrm{max}\sim3.6$\,kpc \citep[e.g.,][]{2013Liu,2014Harrison,2014Husemann,2017Sun}. Alternatively, we estimated the size using the empirical $R_\mathrm{out}-L_\mathrm{[OIII]}$ relation \citep{2018Kang}, 
\begin{equation}\label{eq:OIII-size-lum}
    \log(\frac{R_\mathrm{out}}{\mathrm{kpc}})=(0.28\pm0.04)\log(\frac{L_\mathrm{[OIII]}}{10^{42}\mathrm{erg\,s}^{-1}})+(0.43\pm0.03).
\end{equation}
This yields $R_\mathrm{out}=1.73\pm0.14$\,kpc, consistent with our observational lower limit. Consequently, we adopted $R_\mathrm{out}=1.73^{+1.86}_{-0.58}$\,kpc for our subsequent calculations.

The ionized gas mass of the outflow was calculated from the [OIII] $\lambda5008\AA$ luminosity following Eq.5 in \citet{2015Carniani}, assuming an electron temperature of $T_\mathrm{e}=10^{4}$K and solar metallicity,
\begin{equation}\label{eq:mass-outflow}
    M_\mathrm{out}=0.8M_\odot\left(\frac{L_\mathrm{[OIII]}}{10^{36}\mathrm{erg\,s}^{-1}} \right) \left(\frac{n_\mathrm{e}}{500\mathrm{cm}^{-3}} \right)^{-1}.
\end{equation}
Using the [OIII] outflow luminosity, we derived comparable outflow masses from two spectra, $M_\mathrm{out}=5.8^{+0.7}_{-0.6}\times10^{4}\,M_\odot$ and $M_\mathrm{out}=2.9^{+5.0}_{-2.1}\times10^{4}\,M_\odot$ from AGN and offset-AGN spectra. These consistent masses suggest a relatively uniform distribution of the [OIII] outflow, in align with similar gradients of the electron density and [OIII] luminosity. Assuming a spherical geometry and uniform distribution within a projected distance of $R_\mathrm{sep}\sim2.5$" (the separation between the two pointings), we estimated a lower limit of $M_\mathrm{out}^\mathrm{tot}>A\times M_\mathrm{out}=3.9^{+0.5}_{-0.5}\times10^{5}\,M_\odot$ using the area factor $A=S_\mathrm{circle}/S_\mathrm{fiber}=R_\mathrm{sep}^2/R_\mathrm{fiber}^2\sim6.8$, where $R_\mathrm{fiber}=\mathrm{FWHM}(\mathrm{PSF_\mathrm{g}})/2=0.96$". An upper limit was further estimated by assuming the outflow extends across the entire S\'ersic region ($A=S_\mathrm{S\acute{e}rsic}/S_\mathrm{fiber}\sim65$), yielding $M_\mathrm{out}^\mathrm{tot}<3.8^{+0.5}_{-0.5}\times10^{6}\,M_\odot$. 
Therefore, we adopted a total outflow mass of $M_\mathrm{out}^\mathrm{tot}=0.4\mbox{--}4\times10^{6}\,M_\odot$. 

Under the assumption of a uniform distribution, the mass outflow rate was estimated by dividing the total mass by the corresponding dynamical timescale $\tau_\mathrm{out}$ \citep[within $R_\mathrm{sep}$ or $R_\mathrm{S\acute{e}rsic}$, e.g.,][]{2015Feruglio,2020Lutz}, $\dot{M}_\mathrm{out}=3M_\mathrm{out}/\tau_\mathrm{out}=3M_\mathrm{out}v_\mathrm{out}/R_\mathrm{out}=0.3\mbox{--}1.0\,M_\odot/\mathrm{yr}$, where $\tau_\mathrm{out}\sim v_\mathrm{out}/R_\mathrm{out}\sim3.6\mbox{--}11\,\mathrm{Myr}$. Based on this, the kinetic power and momentum flux are  $\dot{E}_\mathrm{K}=0.5\dot{M}_\mathrm{out}v_\mathrm{out}^2=(1.0\mbox{--}3.1)\times10^{40}\,\mathrm{erg/s}$ and $\dot{P}_\mathrm{out}=\dot{M}_\mathrm{out}v_\mathrm{out}=(0.6\mbox{--}2.0)\times10^{33}\,\mathrm{dyn}$, respectively. 

The estimated kinetic power allows us to exclude stellar winds as a primary driving mechanism, as their contribution is typically several orders of magnitude lower \citep[$\sim10^{34\mbox{--}35}\,\mathrm{erg/s}$; e.g.,][]{2022Commercon}. In contrast, the estimated kinetic power of core-collapse supernovae (SNe), $\dot{E}_K^\mathrm{SN}\approx6.3^{+3.5}_{-3.5}\times10^{40}\mathrm{erg/s}$, based on the scaling relation from \citet{2020Veilleux}, $\sim7\times10^{41}(\mathrm{SFR/M_\odot yr^{-1}})\,\mathrm{erg/s}$, where $\mathrm{SFR}=0.09\,M_\odot\mathrm{yr}^{-1}$ derived in Sect.\ref{subsec:SED}, is comparable with that of the observed outflow. However, we caution that the derived SFR may be overestimated due to the lack of FIR/sub-mm constraints \citep{2022Kim}. Even with the current SFR, an SN-driven scenario remains energetically demanding, as it would require a high coupling efficiency of $\sim15\mbox{--}50\%$ \citep[theoretically $<10\%$;][]{1977Chevalier,2015Walch}. No archival Gamma-ray bursts (GRBs\footnote{https://user-web.icecube.wisc.edu/\~grbweb\_public/index.html}) and SNe \citep{Guillochon2017} were reported in KUG 1208. Alternatively, ultra-luminous X-ray sources (ULXs) are another potential driver due to their comparable power \citep[$\log (L_\mathrm{X}/\mathrm{erg\,s}^{-1})\sim39\mbox{--}41$; e.g.,][]{2023Pinto}, although no ULX in KUG 1208 was reported in the latest ULX catalogue \citep{Walton2022}. Conversely, an AGN origin appears to be a more plausible driver: its bolometric luminosity ($L_\mathrm{bol} \sim 10^{44}\,\mathrm{erg/s}$) is four orders of magnitude higher than the [OIII] outflow kinetic power, and its radiative momentum flux ($L_\mathrm{bol}/c$) is comparable with $\dot{P}_\mathrm{out}$. These energetics favour an AGN-driven origin for the observed ionized outflow.

However, the kinetic power of [OIII] outflows is far below the classical quasar-mode feedback benchmark ($\dot{E}_\mathrm{K}/L_\mathrm{Edd}\sim10^{-6}<0.5\%$). Instead, it is closer to the heating rate required to balance the halo cooling in the context of the radio-mode feedback for early-type galaxies (see Sect.\ref{subsec:UFOs}). In addition, the warm ionized phase typically accounts for only a minor fraction of the total galactic-scale outflow budget, while most of the kinetic energy and mass in kpc-scale winds may be carried by the cold molecular phase \citep[e.g.,][]{2014Cicone, 2017Fiore, 2019Fluetsch}. Therefore, the derived [OIII] energetics represent a conservative lower limit, and the total multi-phase outflow power in KUG 1208 could be substantially higher if a molecular component is taken into account.

Whether outflows can escape the gravitational potential of their host galaxy depends on whether their velocities exceed the local escape velocity $v_\mathrm{esc}$. Following \citet{2019Manzano-King} and \citet{2025Rodriguez}, by assuming a spherical Navarro-Frenk-White \citep[NFW;][]{1997Navarro} dark matter density profile and making use of the abundance matching \citep{2013Moster}, we derived the halo mass from the stellar mass, $M_\mathrm{h}=1.4^{+1.7}_{-0.5}\times10^{12}M_\odot$ \citep[from Eq.12 in][]{2025Rodriguez}. The escape velocity at the center ($r=0$) was obtained as a conservative upper limit, $v_\mathrm{esc} (r=0)=\sqrt{2|\Phi(r=0)|}$, where $\Phi$ is the gravitational potential \citep[see Eq.10 in][]{2001Lokas}. The escape velocity was inferred to be $v_\mathrm{esc}=83^{+26}_{-11}$km/s. Alternatively, the escape velocity could be estimated using the empirical relation with the circular velocity $v_\mathrm{esc}\sim3v_\mathrm{cir}$ \citep{2020Veilleux}, where the circular velocity is $v_\mathrm{cir}=\sqrt{v_\star^2+2\sigma_\star^2}$. The stellar velocity $v_\star$ is negligible in KUG 1208 ($v_\star<10$\,km/s from pPXF), while the stellar dispersion was measured at $\sigma_\star=92\pm5$\,km/s (see Sect.\ref{subsec:ppxf}), yielding $v_\mathrm{esc}\sim390\pm21$\,km/s. This more conservative estimate, comparable to $W_{80}$ velocity, suggests that most kpc-scale [OIII] outflow is unlikely to escape the galactic potential, as expected from its insufficient power to drive AGN feedback.

\section{Discussion}\label{sec:discussion}

We summarize the inferred properties of galaxy and outflows in Tab.\ref{tab:Outflow-comparison} for discussions in this section.

\subsection{Comparison between UFO and [OIII] Outflows}\label{subsec:comparison}

\begin{figure}[htbp]
    \centering        
    \includegraphics[width=\columnwidth]{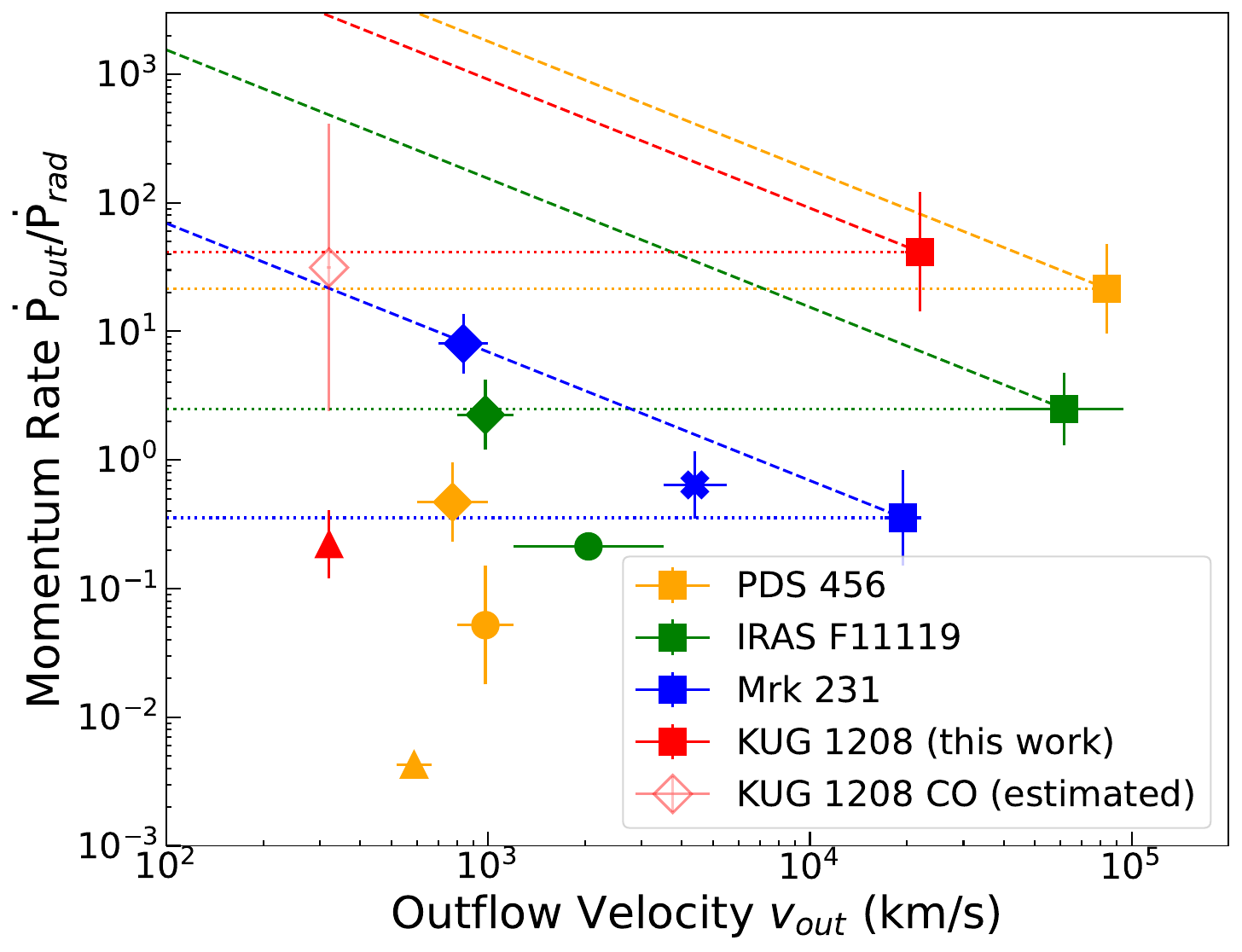}
    \caption{
        Outflow momentum rate, normalized by the AGN radiative momentum ($L_\mathrm{bol}/c$) versus velocity for different gas phases. Data for KUG~1208 (\textit{red}) are compared with three benchmark targets: PDS~456 (\textit{orange}; \citealt{2024Travascio,2025Xrism}), IRAS~F11119+3257 (\textit{green}; \citealt{2018Nardini}), and Mrk~231 (\textit{blue}; \citealt{2015Feruglio}). Markers: UFOs (\textit{squares}), BALs (\textit{crosses}), [NeII-VI] (\textit{circles}), [OIII] (\textit{triangles}), and CO (\textit{diamonds}). \textit{Dotted} and \textit{dashed} lines represent momentum- and energy-conserving theoretical predictions, respectively. 
        The \textit{hollow red diamond} indicates the predicted CO outflow for KUG~1208 (Sec.\ref{subsec:comparison}).
    }
    \label{fig:energetics}
\end{figure}

\begin{table*}
\renewcommand{\arraystretch}{1}
\centering
\caption{Summary of inferred galaxy and outflow properties of KUG 1208+386
}
\begin{tabular}{lcccccccc}
\hline
\hline
\multicolumn{8}{c}{Galaxy}\\
\hline
$M_\star$  & $\sigma_\star$ & $t_\mathrm{quench}$ & SFR  & sSFR & $\log M_\mathrm{BH}$ & $L_\mathrm{bol}$ & $\lambda_\mathrm{Edd}$ \\
   ($10^{10}M_\odot$) & (km/s) & (Gyr) &  ($M_\odot/\mathrm{yr}$) &  ($\mathrm{yr}^{-1}$) &  ($M_\odot$) & ($10^{44}\,\mathrm{erg/s}$) & (\%)\\
\hline
$3.1\pm0.6$ & $92\pm5$ & $\sim3.4$ & $0.09\pm0.05$ & $3\times10^{-12}$ & $7.8\pm0.5$ & $1.3$ & $1.6^{+3.6}_{-1.1}$ \\
\hline
\multicolumn{8}{c}{Outflow}\\
\hline
Phase & Tracer & $M_\mathrm{out}$ & $v_\mathrm{out}$ & $R_\mathrm{out}$ & $\dot{M}_\mathrm{out}$ & $\dot{P}_\mathrm{out}$  & $\dot{E}_\mathrm{out}$\\
 & & ($10^{6}M_\odot$) & (km/s) & (parsec) & ($M_\odot/\mathrm{yr}$)& ($10^{33}$ dyn) & ($10^{40}$ erg/s)\\
\hline
Hot ionized$^\star$ & Fe K & - & $21600^{+8400}_{-6900}$ & $<0.04$ &$0.2\mbox{--}14$ & $24\mbox{--}2150$ & $(0.2\mbox{--}28)\times10^{4}$\\
Warm ionized & [OIII] $\lambda$5008$\AA$ & $0.4\mbox{--}4$ & $315^{+4}_{-4}$ & $1150\mbox{--}3590$ & $0.3\mbox{--}1.0$ & $0.6\mbox{--}2.0$ & $1.0\mbox{--}3.1$\\

\hline
\hline
\vspace{-4mm}
\end{tabular}
\label{tab:Outflow-comparison}
\begin{flushleft}
{$^\star$: Values are inferred at $2\sigma$ uncertainties.}\\
\end{flushleft}
\vspace{-3mm}
\end{table*}
Fig.\ref{fig:energetics} illustrates the wind momentum rate (normalized by the AGN radiative momentum, $L_\mathrm{bol}/c$) as a function of outflow velocities for KUG 1208. Theoretical predictions for momentum-driven and energy-conserving expansion are also overplotted. Three benchmark targets, hosting multiphase winds, ranging from highly ionized UFOs, warmly ionized outflows (BALs, [OIII], or [Ne II-VI]), to cold molecular CO outflows, are also shown: PDS 456 \citep{2019Bischetti,2024Travascio,2025Xrism}, IRAS F11119+3257 \citep[momentum-conserving winds;][]{2018Nardini,2026Seebeck}, and Mrk 231 \citep[energy-conserving winds;][]{2015Feruglio}.

As typically seen in other AGN \citep[e.g.,][]{2017Fiore,2024Travascio,2024Seebeck,2024Lanzuisi}, the warmly ionized outflows in KUG 1208 are less powerful than highly ionized nuclear UFOs and do not align with energy-conserving or momentum-conserving models. This energetic mismatch originates from their inefficient energy transfer, as UFO energy is expected to be rapidly dissipated via radiative cooling at the shock interface \citep{2019Mizumotob}, with the bulk of the momentum ultimately deposited into the cold molecular phase. However, lacking sub-mm observations on KUG 1208 prevents us from directly probing this dominant molecular component.

Despite this limitation, the detected [O III] outflow still provides a valuable tracer of the kpc-scale gas dynamics. To tentatively explore the multi-phase energetics of this system, we attempt to estimate the momentum of the unobserved cold molecular phase using the [O III] outflow as a proxy. Assuming a typical range of $0.1\mbox{--}5\%$ for the fraction of warmly ionized gas momentum relative to the total kpc-scale outflow momentum \citep{2017Fiore,2019Fluetsch}, we estimated the hypothetical momentum rate of the cold molecular gas, shown by the \textit{hollow red diamond} in Fig.\ref{fig:energetics}. If this broad scaling holds true, the multi-phase outflow in KUG 1208 aligns with a momentum-conserving scenario, implying a low overall energy transfer efficiency from the nuclear wind to the host galaxy. Furthermore, under the assumption that the putative cold gas shares the same velocity as the [O III] wind, the total kinetic power could marginally approach the theoretical AGN feedback threshold ($\sim0.5\%L_\mathrm{Edd}$). We caution that applying a population-based scaling relation to an individual source remains speculative. Future sub-mm observations are required to directly measure the energetics of molecular winds and to reliably determine the energy coupling efficiency among the multi-scale outflows.

\subsection{Impacts of Ionized Outflows}\label{subsec:impacts}

Results from both SFH analysis and SED modeling  (Sec.\ref{subsec:ppxf} and \ref{subsec:SED}) identify KUG 1208 as a quiescent galaxy. Although the origin of this quenched state cannot be uniquely determined from the present data alone, AGN feedback is a plausible mechanism for suppressing the star formation \cite[e.g.,][]{2012Fabian}. We therefore examine the energetic feasibility of whether AGN-driven winds can affect the galactic gas reservoir by either expelling or heating the gas.

For a red-sequence galaxy like KUG 1208, the molecular gas fraction $f_\mathrm{gas}\equiv M_\mathrm{gas}/M_\star$ is typically below $1\%$ \citep[e.g.,][]{2017Saintonge,2018Catinella}. Since no direct sub-mm observation is currently available for KUG 1208, we adopted $f_\mathrm{gas}=1\%$ as a conservative upper limit. Together with a stellar mass of $M_\star=3.1\times10^{10}\,M_\odot$, we estimated the cold gas reservoir to be $M_\mathrm{gas}\sim3.1\times10^{8}\,M_\odot$. The corresponding gravitational binding energy is $E_\mathrm{bind}\sim M_\mathrm{gas}\sigma^2_\star=5\times10^{55}\,\mathrm{erg}$, given a stellar dispersion of $\sigma_\star=92$\,km/s. We adopted a characteristic quenching timescale of $\sim3.4$\,Gyr, the interval between the SFR peak and the moment when 90\% of the stellar mass was assembled. Assuming that the currently observed AGN outflows persisted throughout the quenching period with a representative AGN duty cycle of $0.1\%\mbox{--}1\%$ (i.e., 3.4--34\,Myr) for local galaxies \citep[e.g.,][]{2018Aird}, we estimated the cumulative mass loss from [OIII] outflows and UFOs to be $M_\mathrm{out}^\mathrm{[OIII]}\sim(0.1\mbox{--}3.4)\times10^{7}\,M_\odot$ and $M_\mathrm{out}^\mathrm{UFO}\sim(0.2\mbox{--}15)\times10^{7}\,M_\odot$, respectively. The corresponding energy injected into the host galaxy is $E_\mathrm{out}^\mathrm{[OIII]}\sim(0.1\mbox{--}3.4)\times10^{55}\,\mathrm{erg}$ and $E_\mathrm{out}^\mathrm{UFO}\sim(0.08\mbox{--}7.0)\times10^{59}\,\mathrm{erg}$, where the cumulative UFO energy exceeds $E_\mathrm{bind}$ by several orders of magnitude. Assuming a typical galaxy radius of 10\,kpc, the propagation times of the UFO and [OIII] outflows are estimated to be $<0.5$\,Myr and $<31$\,Myr, respectively. Both timescales are significantly shorter than the estimated quenching duration, allowing them to traverse the galaxy and couple with the ISM within the quenching period.

Consequently, while [OIII] winds may lack the kinetic power to drive large-scale feedback, UFOs are energetically capable of affecting the gas reservoir. Despite their transient nature \citep[e.g.,][]{2018Reeves,2025Gu}, recurrent UFOs covering as low as $1\%$ of the AGN lifetime remain energetic enough to effectively impact the host galaxy. In fact, the high UFO detection rate in AGN \citep[$\sim30\mbox{--}40\%$;][]{2010Tombesi,2021Chartas,2023Matzeu} further supports the possibility that such intermittent winds may play an important role in maintaining the quiescent state of KUG 1208.

\subsection{Comparison with Other Galaxies}\label{subsec:comparison}

\begin{figure}[htbp]
    \centering        
    \includegraphics[width=\columnwidth]{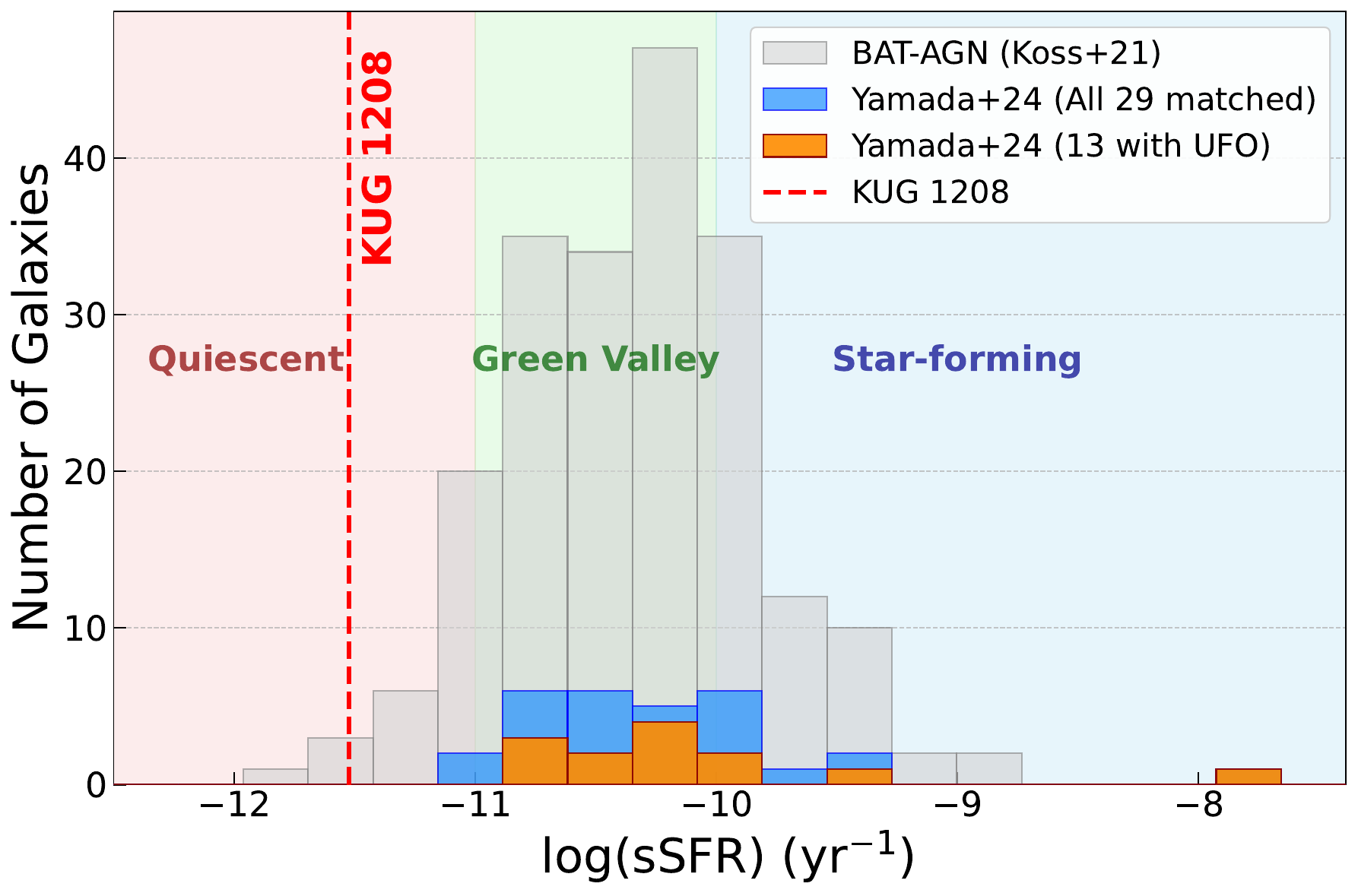}
    \caption{sSFR distributions of UFO-host versus non-UFO galaxies. The underlying \textit{gray} histogram represents the parent sample of 213 local Seyfert 1 galaxies from the BAT-AGN catalog \citep{Koss2021}. The \textit{blue} and \textit{orange} overlapping histograms display the cross-matched 29 sources from the \citet{2024Yamada} sample and its subset of 13 UFO-detected sources, respectively. KUG 1208 is marked by the \textit{vertical red dashed} line at $\log(\text{sSFR}) \approx -11.5 \text{ yr}^{-1}$. The \textit{shaded red, green, and blue} regions denote the typical quiescent, GV, and SF galaxy phases, respectively \citep[e.g.,][]{Wetzel2012}. 
    }
    \label{fig:comparison}
\end{figure}

While UFOs are widely detected in AGN, KUG 1208 represents \textit{the first reported case of UFO detection in a quiescent host}. We demonstrate the uniqueness of KUG 1208 by comparing its sSFR with those of other local AGN hosts. As illustrated in Fig.~\ref{fig:comparison}, we present the sSFR distribution of a parent sample comprising 213 local hard X-ray-selected galaxies from the BAT-AGN catalog \citep{Koss2021}. The sSFRs of these galaxies were directly derived from sub-mm observations of their molecular gas. We cross-matched this parent catalog with the legacy survey of X-ray outflows in 132 AGN from \citet{2024Yamada}, yielding a total of 29 matched sources, 13 of which host confirmed UFOs. The distribution clearly demonstrates that the UFO-detected AGN galaxies predominantly reside within the SF main sequence or the transitioning GV regime ($\log(\text{sSFR}/\text{yr}^{-1}) \gtrsim -11$). 

In contrast, KUG 1208 exhibits an exceptionally low sSFR ($\log(\text{sSFR}/\text{yr}^{-1}) \approx -11.5$), placing it in the quiescent regime. This severe environmental mismatch, a powerful AGN wind operating within a long-dead host, fundamentally distinguishes KUG 1208 from the canonical population of UFO host galaxies. This finding directly challenges the prevailing paradigm that UFOs are predominantly hosted by gas-rich galaxies \citep{1998Schinnerer,2003Pounds,2009Riechers,2013Veilleux,2015Tombesi,2018Braito,2018Sanfrutos, 2024Travascio}. It implies that UFO triggering is not strictly coupled to the global star-forming status of the host, but is instead governed by the local circumnuclear environment. This is supported by the observed high column densities of both scattering and LOS host materials ($\log (N_\mathrm{H}/\mathrm{cm}^{-2})\gtrsim23$; Sect.\ref{subsec:model-B}), which indicates a localized reservoir capable of fueling the central SMBH.

Integrating these findings into the broader framework of galaxy evolution \citep{2014Heckman,2018Fischer}, AGN feedback in KUG 1208 likely operates in a `maintenance' mode, \textit{dominated by AGN winds instead of radio jets} \citep[e.g.,][]{2012Fabian,2013Gaspari,2014Heckman}. Although global star formation ceased $\sim 9$\,Gyr ago, the presence of an AGN indicates that the nucleus is not entirely dormant but rather undergoes intermittent re-triggering. Instead of driving large-scale gas ejection, AGN winds likely sustain the quiescence by injecting mechanical energy into the gas reservoir via large-scale turbulence. This scenario is further supported by our finding that the bulk of the [OIII] outflow remains gravitationally bound, implying that the wind energy is dissipated within the host galaxy. Such wind-driven turbulence can effectively prevent gas from cooling and collapsing, thereby inhibiting star formation. This `maintenance' scenario provides a self-consistent explanation for the coexistence of powerful feedback and dense nuclear gas in a galaxy that has remained dormant since the early universe.

\section{Conclusions}\label{sec:conclusions}
We present a multi-wavelength study of the Seyfert 1.2 galaxy KUG 1208+386, combining X-ray (\textit{XMM-Newton}+\textit{NuSTAR}) and optical (\textit{DESI}) data to probe the coupling between nuclear and galactic outflows. Our findings are summarized as follows:
\begin{itemize}
    \item \textbf{First UFO detection in a quiescent host}: We report the first detection of a highly ionized UFO ($v_\mathrm{UFO}\sim0.07c$) in a massive galaxy that has been quenched for $\sim$9\,Gyr. This discovery challenges the convention that UFOs are exclusive to gas-rich, star-forming systems. 
    \item \textbf{Circumnuclear triggering}: The high column densities ($\log (N_\mathrm{H}/\mathrm{cm}^{-2}) \gtrsim 23$) from X-ray spectroscopy reveal a dense, localized gas reservoir fueling the SMBH. This implies that UFO triggering is decoupled from global host properties and is instead governed by the nuclear environment.
    \item \textbf{A possible momentum-conserving outflow}: The [OIII] outflow extends to $\gtrsim1.2$\,kpc despite its modest energetics, in contrast to the powerful nuclear UFO. Comparing the UFO, ionized [OIII] outflow, and inferred cold molecular gas indicates a momentum-conserving outflow in KUG 1208.
    \item \textbf{Wind-driven maintenance mode of AGN feedback}: The coexistence of a powerful UFO and a local dense reservoir in a quenched host suggests a `maintenance' feedback mode in KUG 1208. Intermittent wind-driven turbulence, rather than radio jets, suffices to suppress gas cooling and sustain quiescence without requiring bulk gas expulsion.
\end{itemize}

Future deep high-resolution X-ray (\textit{XMM-Newton} and \textit{XRISM}) and sub-mm observations will be crucial to constrain the multi-phase outflow coupling and provide a definitive validation of this wind-driven maintenance-mode scenario.

\begin{acknowledgements}
This work is supported by the MCIU program Unidad de Excelencia María de Maeztu CEX2020-001058-M. M.S. acknowledges support by the State Research Agency of the Spanish Ministry of Science and Innovation under the grants 'Galaxy Evolution with Artificial Intelligence' (PGC2018-100852-A-I00) and 'BASALT' (PID2021-126838NB-I00) and the Polish National Agency for Academic Exchange (Bekker grant BPN/BEK/2021/1/00298/DEC/1). This work was partially supported by the European Union's Horizon 2020 Research and Innovation program under the Maria Sklodowska-Curie grant agreement (No. 754510). VRM and M.M. acknowledge support from the Spanish Ministry of Science and Innovation through the project PID2021-124243NBC22 and PRE2022-104649, respectively. CP is supported by INAF Large Grant 2023 BLOSSOM O.F. 1.05.23.01.13.

\end{acknowledgements}

\bibliographystyle{aa}
\bibliography{ref.bib}

\begin{appendix}

\onecolumn
\section{X-ray Data and Analysis}\label{app:sec:raw_data}

\begin{figure}[htbp]
    \centering        
    \includegraphics[width=\columnwidth]{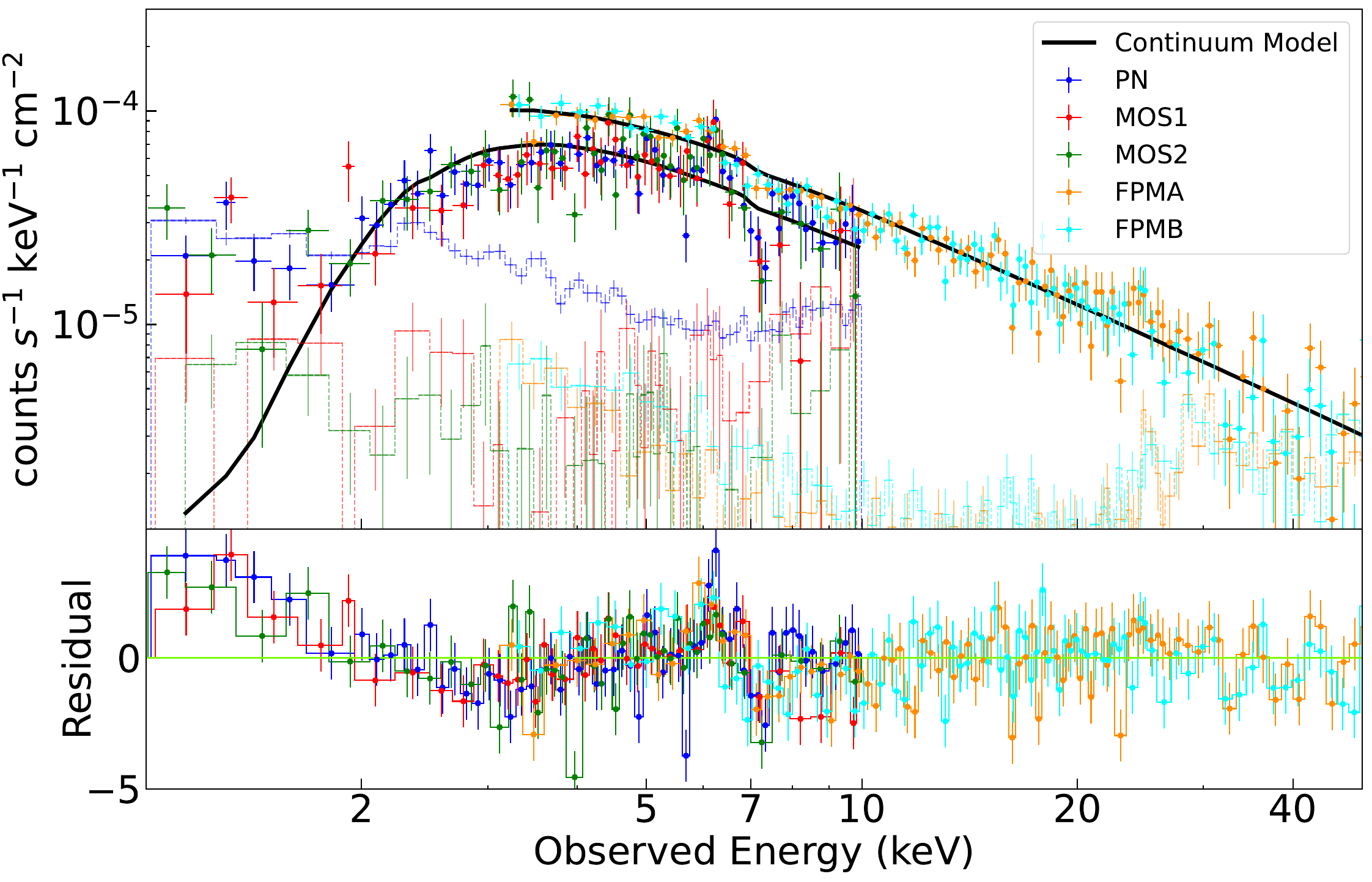}
    \caption{X-ray flux-energy (\textit{upper}) spectra from EPIC-pn (\textit{blue}), EPIC-MOS1 (\textit{red}), EPIC-MOS2 (\textit{green}), FPMA (\textit{orange}), and FPMB (\textit{cyan}), shown together with their corresponding background spectra (faint, diluted points in matching colors). Spectra were rebinned for clarity. The residual spectra against the continuum model (an absorbed Comptonization) are shown in the \textit{lower} panel.}
    \label{app:fig:raw_data}
\end{figure}

\begin{figure}[htbp]
    \centering        
    \includegraphics[width=0.7\columnwidth]{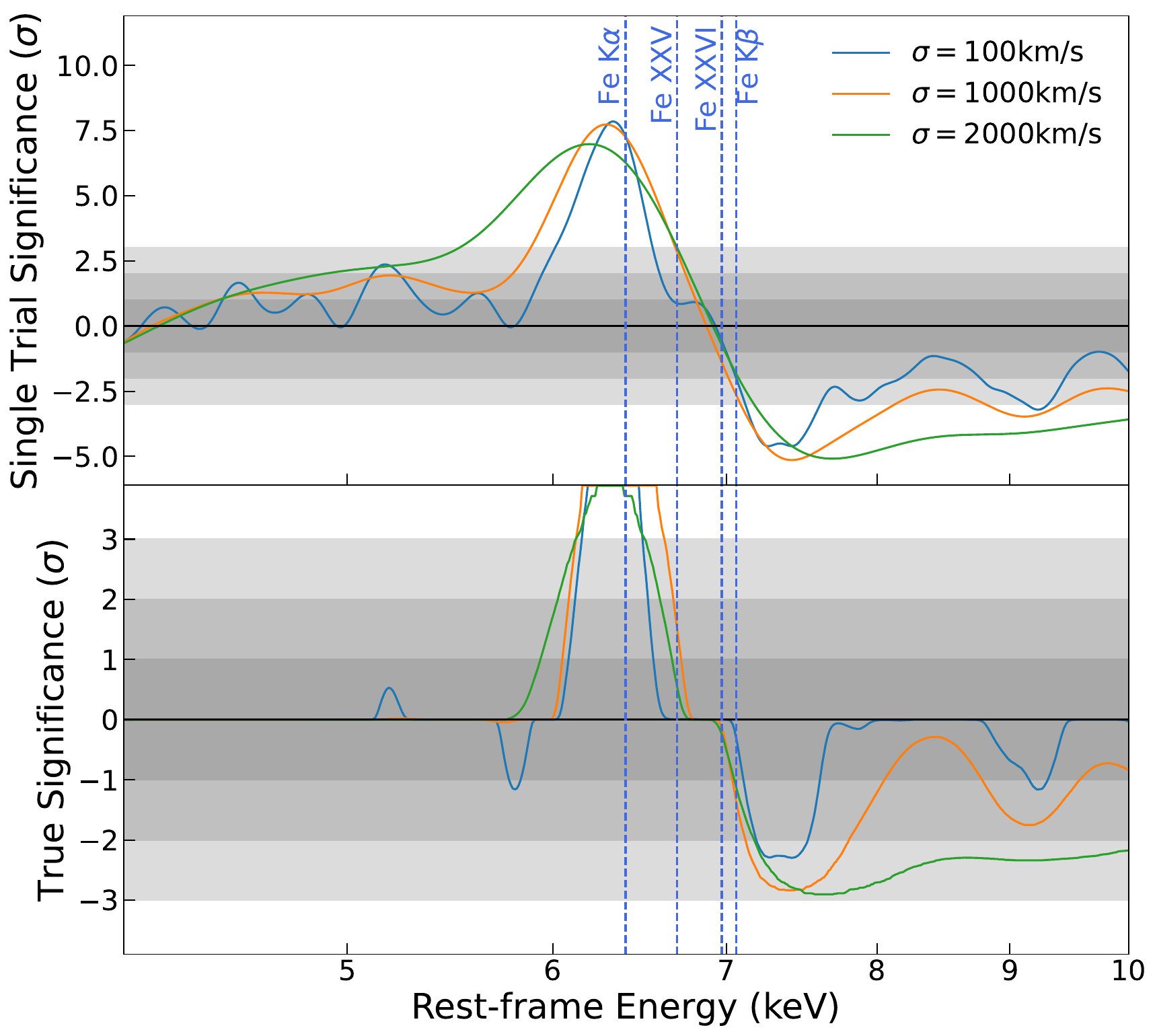}
    \caption{The single trial (\textit{top}) and true (\textit{bottom}) significance of line features with different Gaussian widths over the $4\mbox{--}10$\,keV band. The dark grey, grey, and light grey regions correspond to the significance of $1\sigma, 2\sigma$, and $3\sigma$, respectively. The laboratory positions of Fe K$\alpha$, XXV, XXVI, and K$\beta$ are marked by the vertical dashed lines.}
    \label{app:fig:line}
\end{figure}

To estimate the line significance, we performed two Gaussian line search methods on model A.
Briefly, the first method is a sliding Gaussian line scan applied directly to the observed spectrum. We added a Gaussian component to model A and stepped its centroid energy through the $2\mbox{--}10$\,keV energy band. For each trial, both the line energy and width were fixed, while the normalization was allowed to vary. The improved statistics were recorded at each trial energy bin and width. The square root of the $\Delta$C-stat ($\sqrt{\Delta\mathrm{C\mbox{--}stat}}$) can serve as a rough proxy for the line significance \citep{1979Cash}. 

This method is efficient but only provides the single-trial significance without considering the look-elsewhere (LE) effect, where the significant signals without known energies may originate from the noise. The second method is the cross-correlation-based MC Gaussian line search accounting for the LE effect \citep[e.g.,][]{2021Kosec,2022Xu,2024Xu}. Classically, one would simulate spectra from the best-fitting continuum model, repeat the Gaussian line scan on each simulated spectrum, and record the strongest spurious feature found anywhere in the searched band. The global significance of the observed feature is then obtained by comparing its strength with the distribution of these maximum noise-induced features.
The difference between our approach and the classical MC method \citep[e.g.,][]{2010Tombesi} is that instead of performing the spectral fitting with C-stat (or $\chi^2$-stat) , this method adopts the cross-correlation (CC) between the Gaussian model spectra and residual spectra against the continuum model, which speeds up the computational efficiency by a factor of $10^4$. 
The true significance is derived from the comparison between the observed line and any significant features over the entire simulated spectra, not limited to the same energy bin, in case of random shifts of UFOs.

For both methods, we tried various Gaussian line widths from $100\mbox{--}2000$\,km/s and took a step of 0.01\,keV among the $2\mbox{--}10$\,keV band. 10000 MC simulations were performed in the second approach, corresponding to the upper limit of $3.9\sigma$. Results are shown in Fig.\ref{app:fig:line}, where in both methods, Fe-K emission is robustly detected at $>3.9\sigma$, and an absorption line is located at $7.25^{+0.07}_{-0.07}$\,keV with a significance of $\gtrsim3\sigma$. The absorption feature below 6 keV is present only in the EPIC-pn spectrum and is therefore attributed to an instrumental or statistical artifact. 
\begin{figure}[h]
    \centering        
    \includegraphics[width=\columnwidth]{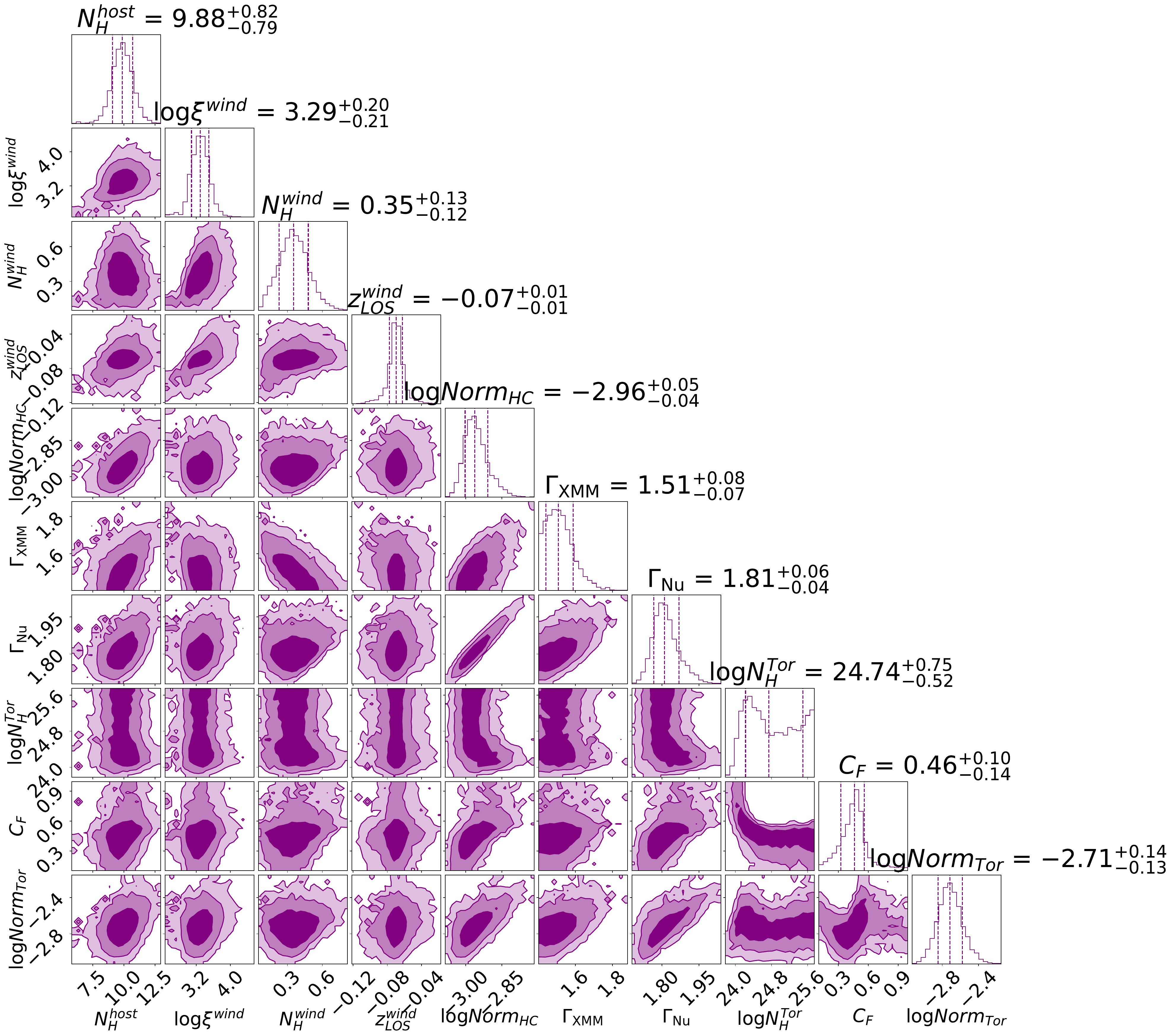}
    \caption{The normalized probability distributions of the posteriors of the best-fit model (Model C: Comptonization+Torus+Wind). The units of the host galactic absorption $N_{H}^{host}$ and the wind column density $N_{H}^{wind}$ are $10^{22}\,\mathrm{cm}^{-2}$ and $10^{24}\,\mathrm{cm}^{-2}$, respectively. Contours represent the $1\sigma$, $2\sigma$, and $3\sigma$ confidence intervals.}
    \vspace{-4mm}
    \label{app:fig:corner}
\end{figure}
\clearpage
\section{Spectral Energy Distribution}\label{app:sec:SED}

\begin{figure}[htbp]
    \centering        
    \includegraphics[width=0.49\columnwidth]{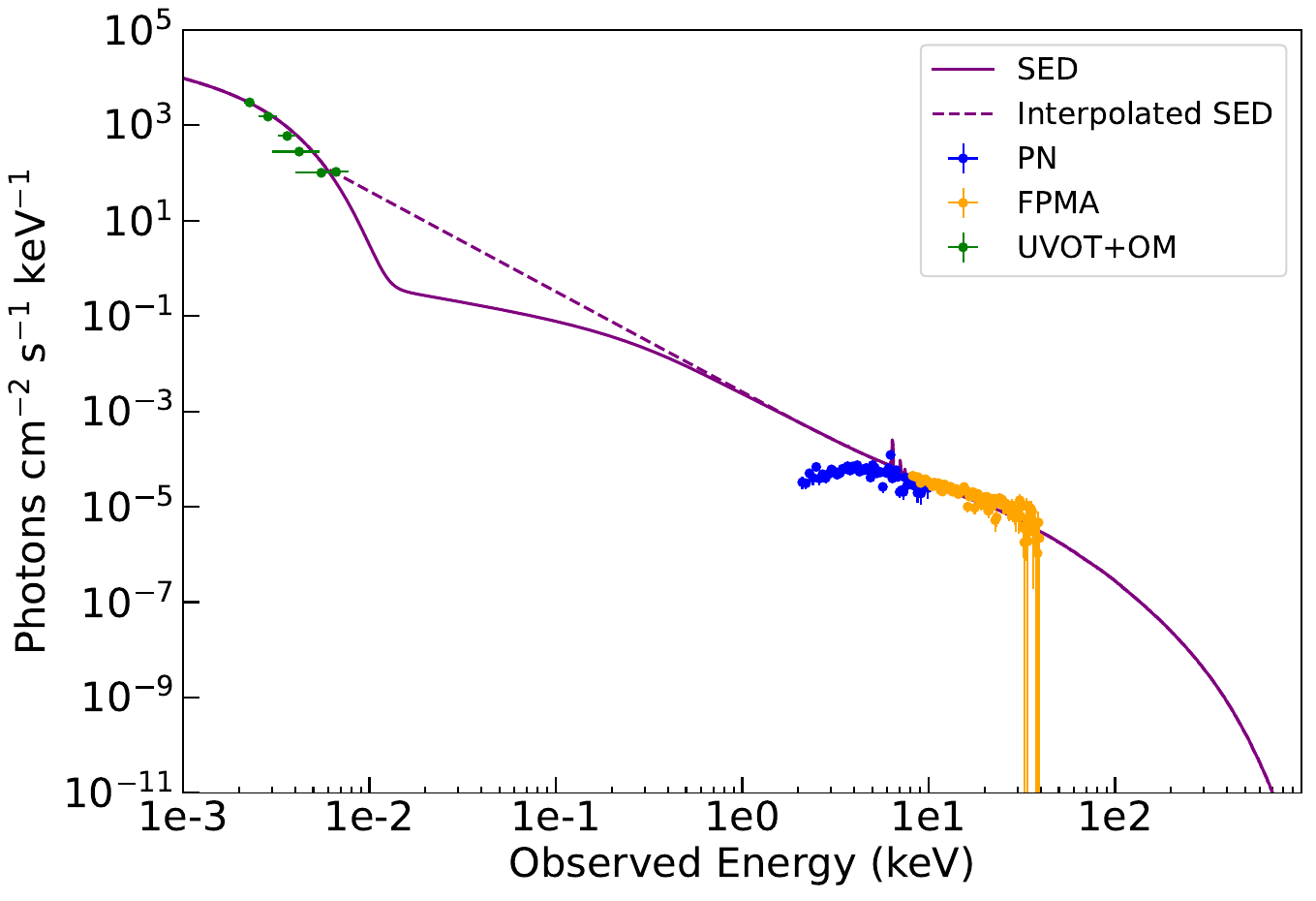}
    \caption{Intrinsic unabsorbed SEDs of KUG 1208 generated by incorporating a disk blackbody with a temperature of $\sim1$\,eV into Model B, overlaid with the EPIC-pn (\textit{blue}), FPMA (\textit{orange}), OM+UVOT data (\textit{green}). EPIC-MOS1/2 and FPMB data are not shown for clarity. The fitted and interpolated (between $6$\,eV and 2\,keV) SEDs are shown by the \textit{purple solid} and \textit{dashed} lines, respectively.  
    }
    \label{app:fig:SED}
\end{figure}

In AGN, UV and a large portion of optical emission primarily originate from the accretion disk. To constrain the intrinsic SED (Fig.\ref{app:fig:SED}), we incorporated the UV and optical photometry data alongside a disk blackbody component (\texttt{diskbb}) into Model B, yielding a fitted temperature of $\sim1$\,eV. Both Galactic and host-galaxy dust extinction were accounted for using the \texttt{redden} model, adopting $\mathrm{E(B-V)}=0.02$ \citep{2011Schlafly} and $\mathrm{E(B-V)}=0.23$ (derived in Sec.\ref{subsec:optical-line-spectra}), respectively. It yielded an ionizing luminosity ($1\mbox{--}1000$\,Ryd) of $L_\mathrm{ion}\sim 2.4\times10^{43}\,\mathrm{erg/s}$ and a bolometric luminosity ($1\,\mathrm{eV}\mbox{--}1000\,\mathrm{keV}$) of $L_\mathrm{bol}\sim1.5\times10^{44}\,\mathrm{erg/s}$, which is comparable with with the value inferred from the \textit{Swift}/BAT data \citep[$L_\mathrm{bol}=1.6\times10^{44}\,\mathrm{erg/s}$,][]{2022Koss}.

Furthermore, to assess the impact of a missing soft X-ray excess in our modeling, we conservatively interpolated between the FUV (UVW2: $1928\AA$) and hard X-ray (2\,keV) bands (\textit{dashed} line in Fig.\ref{app:fig:SED}). It produced a ionizing luminosity of $L_\mathrm{ion}\sim 4.7\times10^{43}\,\mathrm{erg/s}$, and a bolometric luminosity of $L_\mathrm{bol}\sim2.5\times10^{44}\,\mathrm{erg/s}$. Even with these increases, the resulting Eddington ratio (derived in Sec.\ref{subsubsec:BH-mass}) remains within the typical Seyfert regime. The derived UFO properties remain unchanged, except for a marginal increase in the ionization parameter ($\Delta\log\xi=0.15$) that is within the uncertainties. Since these differences do not alter our main conclusions, we retained the fitted SED for subsequent analysis.

\clearpage
\newpage
\section{PyQSOFit Results of DESI Spectra}\label{app:sec:pyQSOFit}

\begin{table}[htbp]
\renewcommand{\arraystretch}{1.1}
\setlength{\tabcolsep}{2pt} %
\centering
\caption{Best-fit properties of lines of interest in KUG1208 from \textit{DESI} spectra fitting.}
\label{app:tab:lines_complex}
\begin{tabular}{lccccc}
\hline
\hline
Line & Wave & Comp$^\star$ & Dispersion &  $F^\dagger$ ($10^{-15}$) & $L^\ddagger$ ($10^{40}$) \\
     & ($\text{\AA}$) & & $\sigma$ (km/s) & ($\mathrm{erg/s/cm^2}$) & (erg/s) \\
\hline
\multicolumn{6}{c}{{AGN Spectrum}} \\
\hline
\multirow{4}{*}{H$\beta$} & 4862.8 & N & $106\pm1^{a}$ & $7.7\pm0.1$ & $2.0\pm0.1$ \\
         & 4857.5 & I & $553\pm6^b$ & $9.6\pm0.4$ & $2.5\pm0.2$ \\ 
         & 4874.8 & I & $899\pm8^b$ & $1.7\pm0.4$ & $0.4\pm0.1$ \\ 
         & 4862.3 & B & $2885\pm12^c$ & $51.4\pm1.0$ & $13.3\pm0.8$ \\ 
\hline
\multirow{2}{*}{[OIII]} & 4960.4 & N & $106^{a}$ & $15.8\pm0.2$ & $4.0\pm0.2$ \\ 
 & 4960.1 & O & $240\pm4^{d}$ & $11.9\pm0.4$ & $3.0\pm0.2$ \\ 
\hline
\multirow{2}{*}{[OIII]} & 5008.3 & N & $106^{a}$ & $47.9\pm0.7$ & $12.1\pm0.7$ \\ 
 & 5008.1 & O & $240^{d}$ & $36.0\pm1.3$ & $9.1\pm0.6$ \\ 
\hline
\multirow{4}{*}{H$\alpha$} & 6564.7 & N & $106^a$ & $27.6\pm0.3$ & $5.7\pm0.2$ \\ 
 & 6557.6 & I & $553^b$ & $77\pm2$ & $15.9\pm0.7$ \\ 
 & 6581.0 & I & $899^b$ & $47\pm2$ & $9.7\pm0.6$ \\ 
 & 6564.0 & B & $2885^c$ & $262\pm2$ & $54\pm2$ \\ 
\hline
[SII] & 6718.4 & N & $106^{a}$ & $6.4\pm0.1$ & $1.30\pm0.05$ \\ 
{[SII]} & 6732.8 & N & $106^{a}$ & $6.2\pm0.1$ & $1.26\pm0.05$ \\ 
\hline
\multicolumn{6}{c}{{Offset-AGN Spectrum}} \\
\hline
\multirow{4}{*}{H$\beta$} & 4862.7 & N & $85\pm3^{e}$ & $0.58\pm0.06$ & $0.15\pm0.06$ \\
         & 4856.2 & I & $546\pm21^f$ & $0.5\pm0.2$ & $0.13\pm0.07$ \\ 
         & 4876.7 & I & $890\pm20^f$ & $0.06\pm0.13$ & $0.01\pm0.03$ \\ 
         & 4862.9 & B & $3020\pm87^g$ & $5.2\pm0.4$ & $1.4\pm0.5$ \\ 
\hline
\multirow{2}{*}{[OIII]} & 4960.3 & N & $85^{e}$ & $0.86\pm0.08$ & $0.23\pm0.09$ \\ 
 & 4959.7 & O & $187\pm11^{h}$ & $1.2\pm0.2$ & $0.31\pm0.13$ \\ 
\hline
\multirow{2}{*}{[OIII]} & 5008.2 & N & $85^{e}$ & $2.6\pm0.2$ & $0.7\pm0.3$ \\ 
 & 5007.7 & O & $187^{h}$ & $3.7\pm0.6$ & $1.0\pm0.4$ \\ 
\hline
\multirow{4}{*}{H$\alpha$} & 6564.6 & N & $85^e$ & $2.1\pm0.1$ & $0.44\pm0.12$ \\ 
 & 6555.9 & I & $546^f$ & $5.5\pm0.3$ & $1.2\pm0.3$ \\ 
 & 6583.6 & I & $890^f$ & $3.1\pm0.4$ & $0.6\pm0.2$ \\ 
 & 6564.9 & B & $3020^g$ & $20\pm1$ & $4.2\pm1.2$ \\ 
\hline
[SII] & 6718.3 & N & $85^{e}$ & $0.95\pm0.05$ & $0.19\pm0.11$ \\ 
{[SII]} & 6732.6 & N & $85^{e}$ & $0.71\pm0.04$ & $0.15\pm0.01$ \\ 
\hline
\hline
\end{tabular}
\begin{flushleft}
\footnotesize
\textit{Notes.} $\star$: N: Narrow; I: Intermediate; B: Broad; O: Outflow. \\
$\dagger$: Line flux after Galactic extinction correction. \\
$\ddagger$: Line luminosity after Galactic+Host extinction correction using the narrow Balmer decrement. \\
${a}\mbox{--}{h}$: Dispersions of lines within the same component are linked.
\end{flushleft}
\vspace{-5mm}
\end{table}
%

\twocolumn
\begin{figure*}[htbp]
    \centering        
    \includegraphics[width=0.85\textwidth]{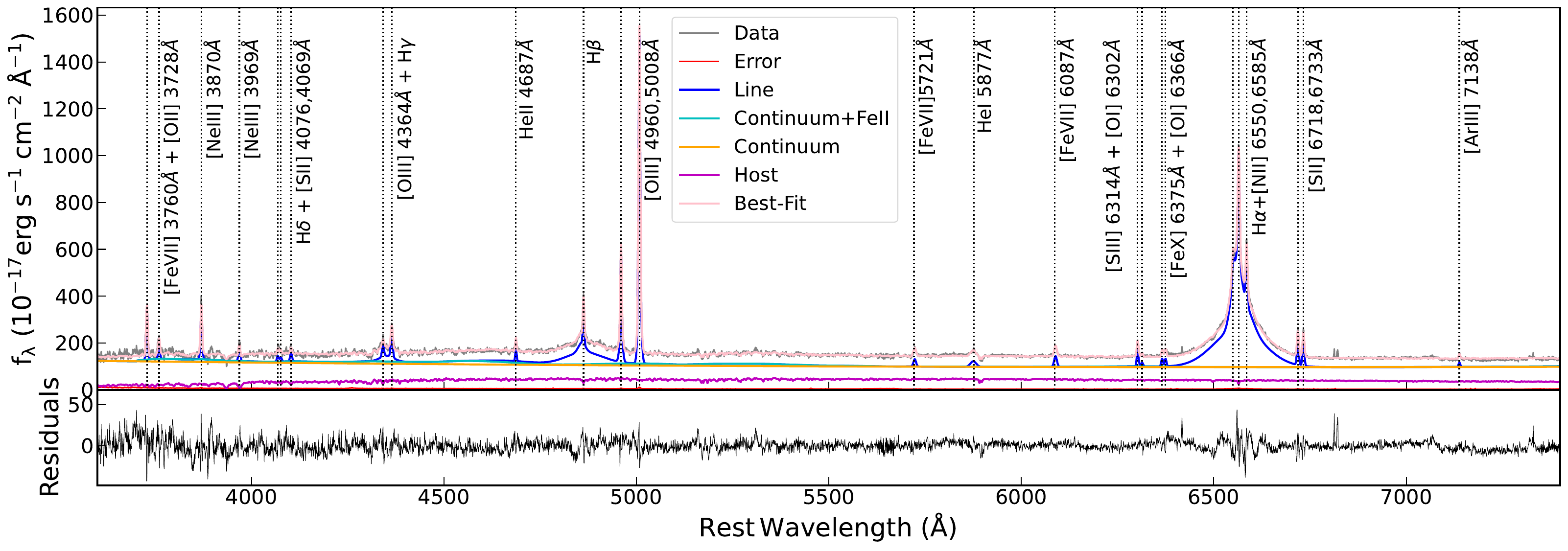}
    \includegraphics[width=0.85\columnwidth]{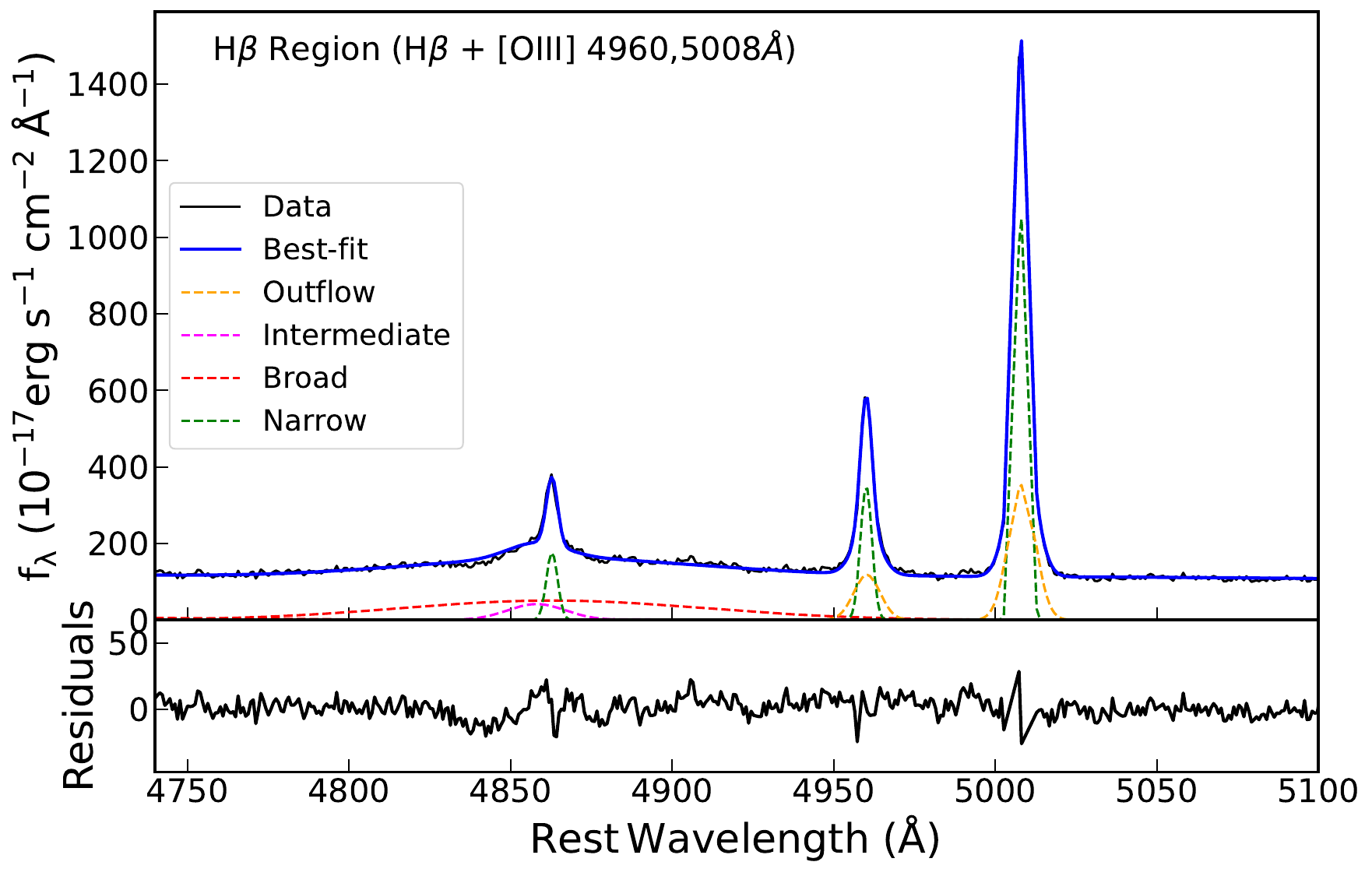}
    \includegraphics[width=0.85\columnwidth]{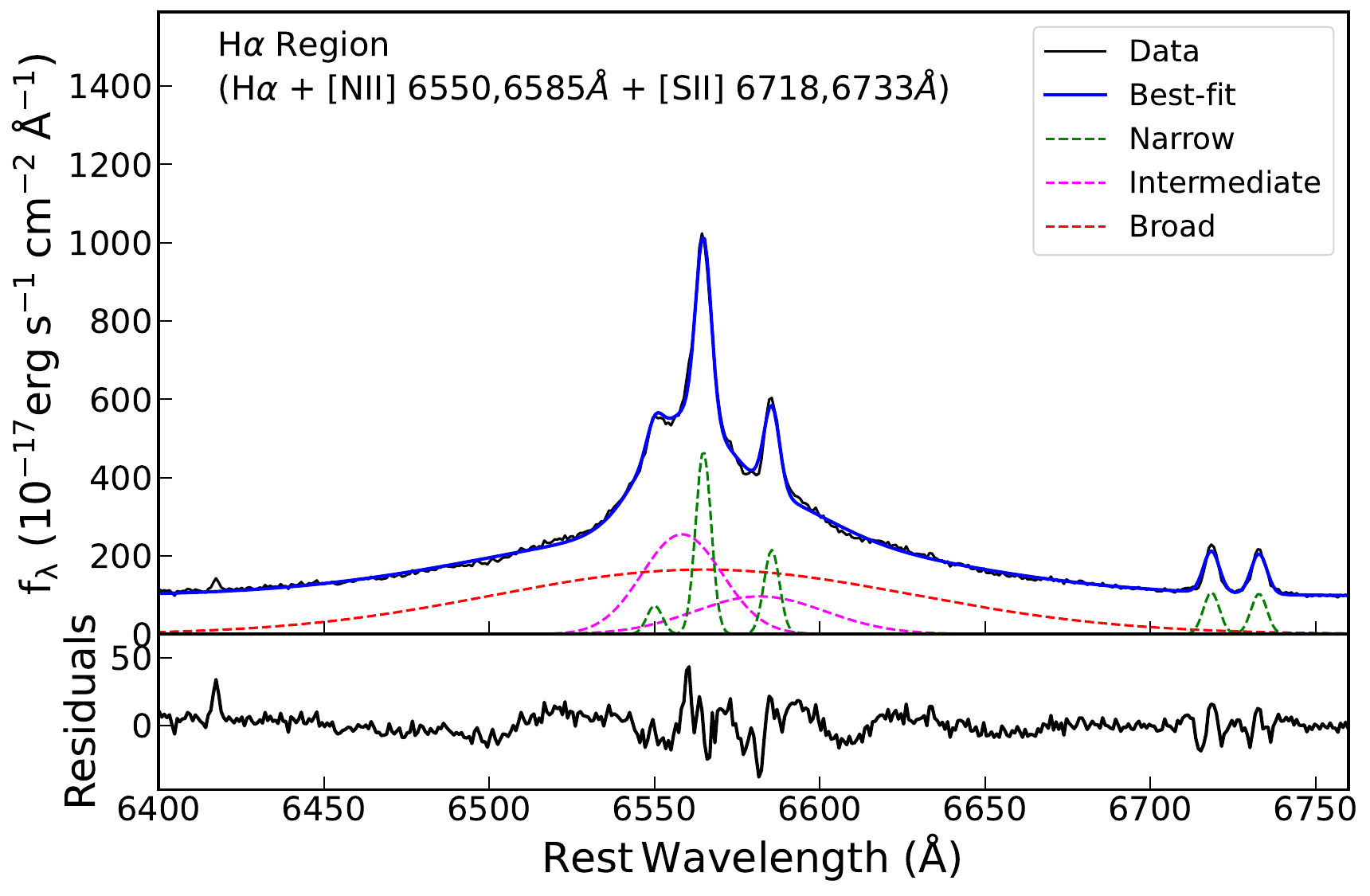}
    \caption{\textit{Upper}: \textit{DESI} AGN spectrum of KUG 1208 (\textit{top}) and the corresponding residuals (\textit{bottom}). The \textit{gray} and \textit{red} curves show the observed spectrum and its uncertainty. The \textit{blue}, \textit{cyan}, \textit{yellow}, \textit{purple}, and \textit{pink} curves represent the fitted emission lines, AGN continuum plus [FeII] emission, AGN continuum, host galaxy stellar emission, and best-fitting total model, respectively. The residuals are defined as $\mathrm{data}-\mathrm{model}$. Fitted emission lines are marked by vertical dashed lines. 
    \textit{Lower}: Zoom-in on the H$\beta$ (\textit{Left}) and H$\alpha$ (\textit{Right}) regions. Lines are fitted by several components, where narrow, outflow, intermediate and broad components are marked by \textit{green}, \textit{orange}, \textit{magenta}, and \textit{red dashed} lines, respectively. 
    }
    \label{fig:DESI-backup}
\end{figure*}

\begin{figure*}[htbp]
    \centering        
    \includegraphics[width=0.85\textwidth]{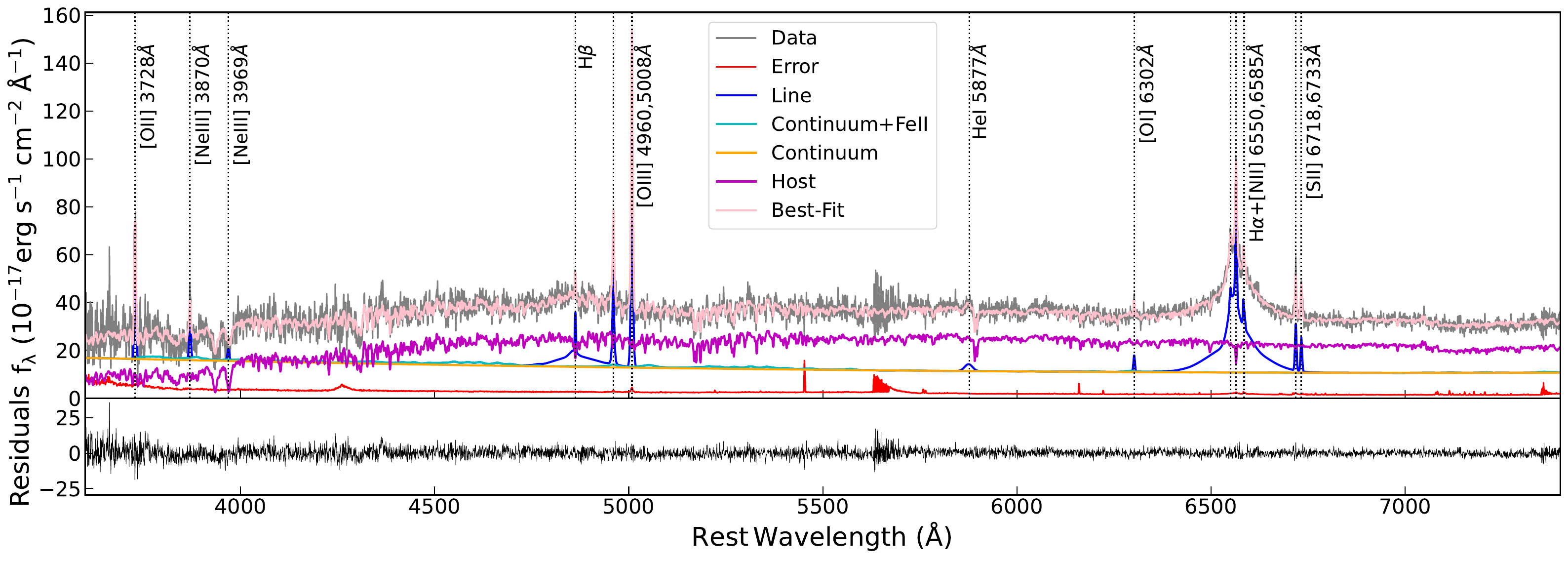}
    \includegraphics[width=0.85\columnwidth]{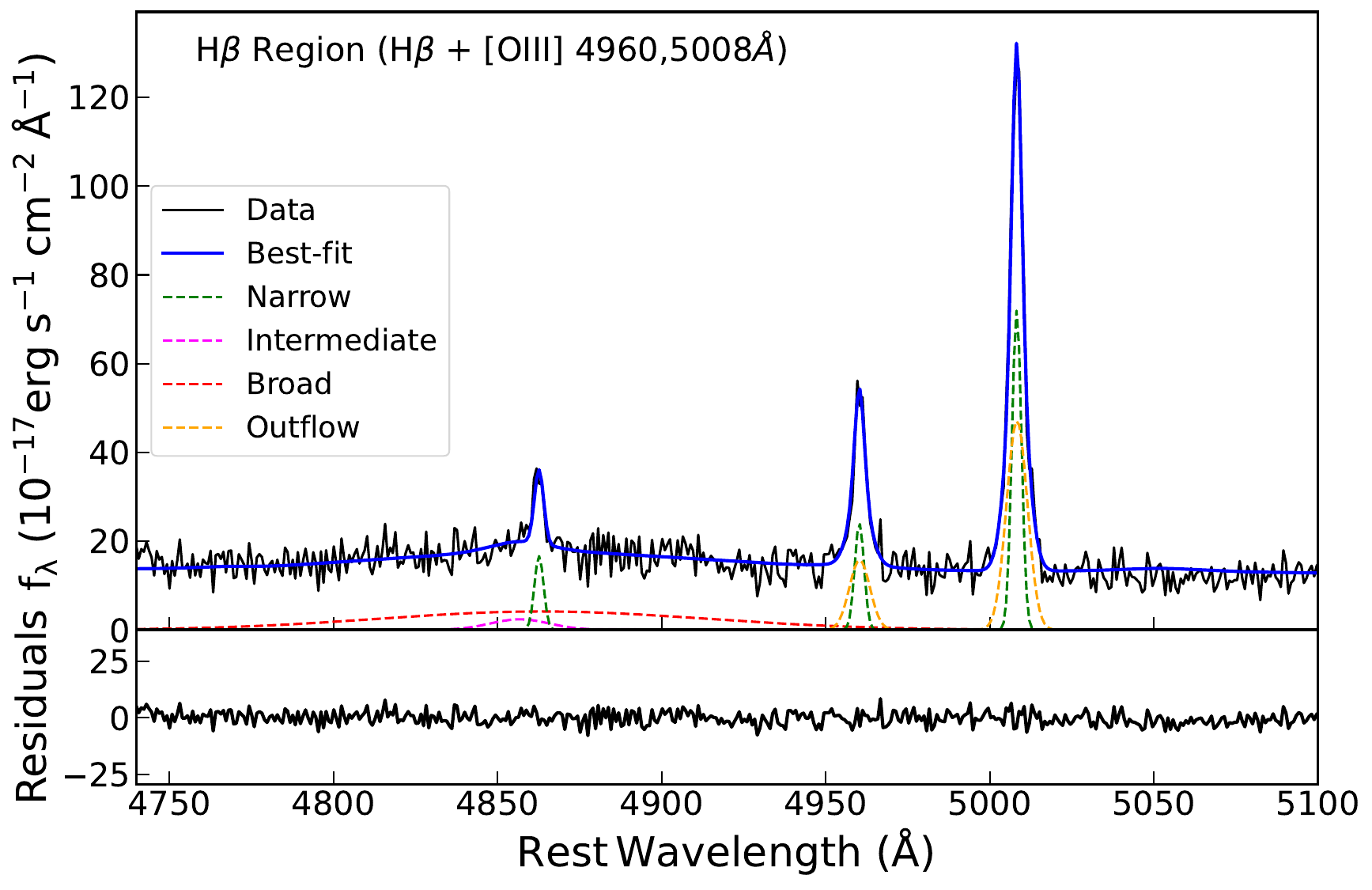}
    \includegraphics[width=0.85\columnwidth]{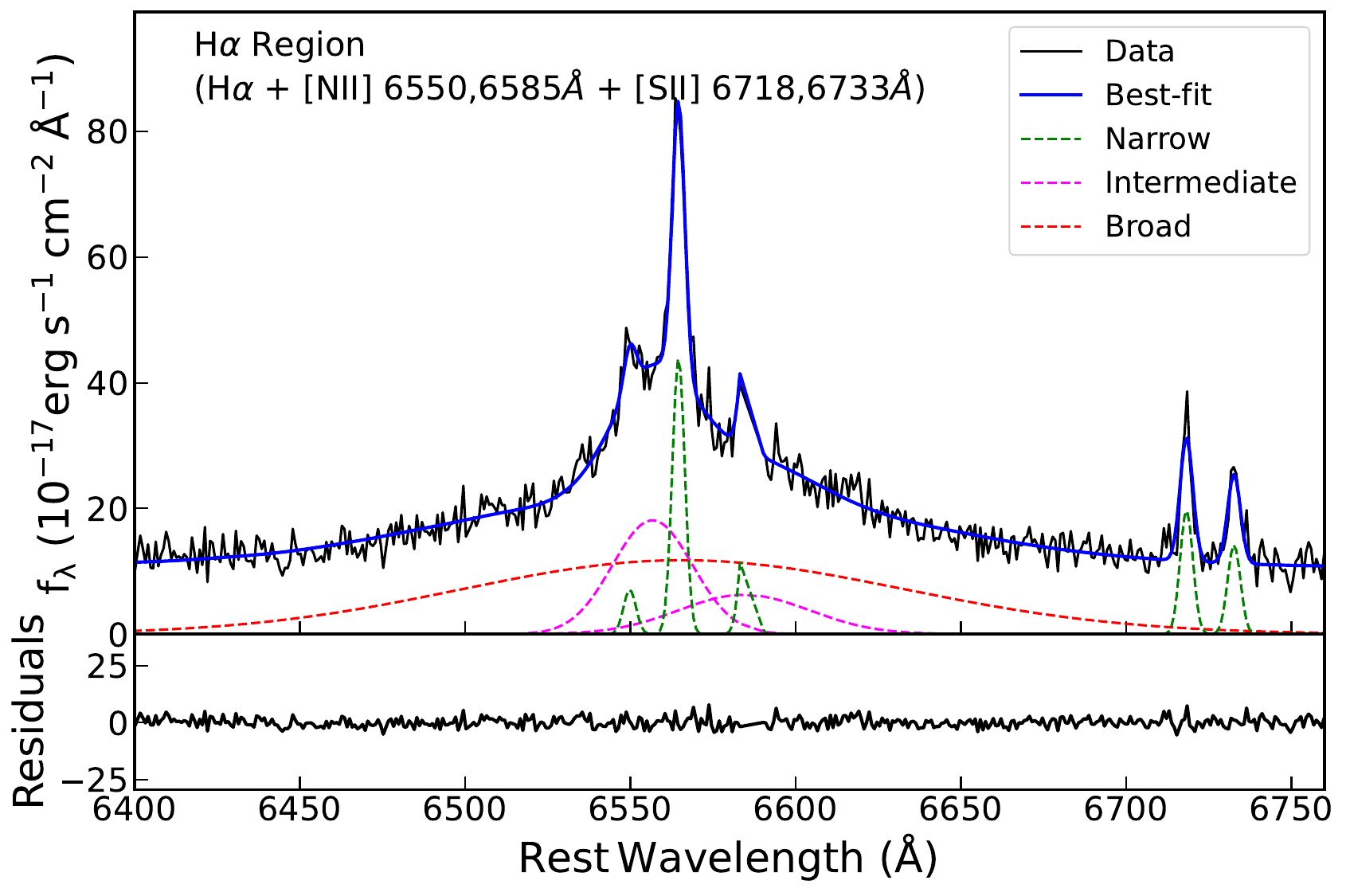}
    \caption{\textit{DESI} offset-AGN spectrum of KUG 1208 (\textit{top}) and the corresponding residuals (\textit{bottom}). Similar to Fig.\ref{fig:DESI-backup} with no coronal lines. 
    }
    \label{app:fig:DESI-bright}
\end{figure*}

\clearpage
\newpage
\onecolumn
\section{Host Galaxy Analysis}\label{app:sec:host}

\begin{figure}[h]
    \centering        
    \includegraphics[width=0.49\columnwidth]{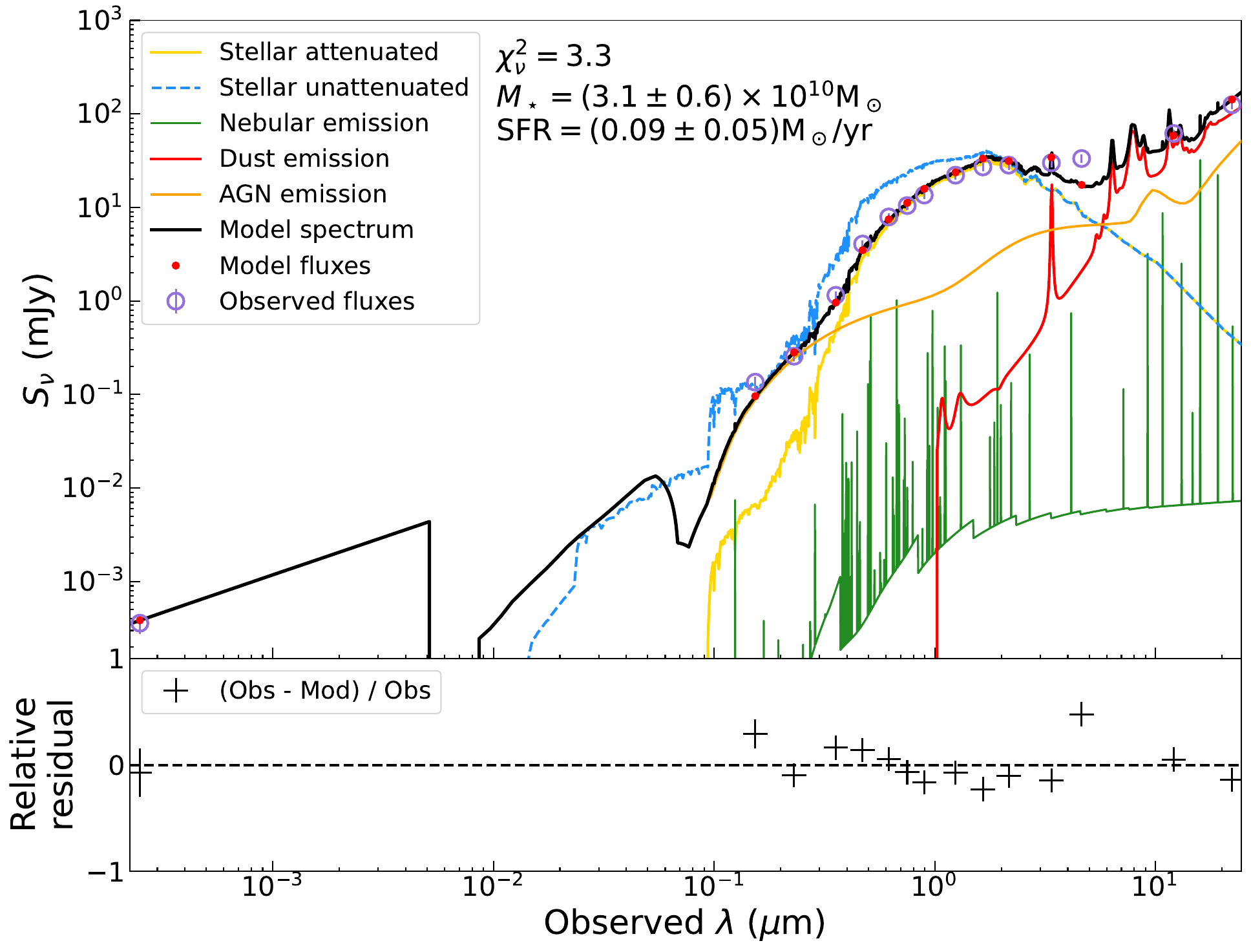}
    \includegraphics[width=0.49\columnwidth,height=6cm]{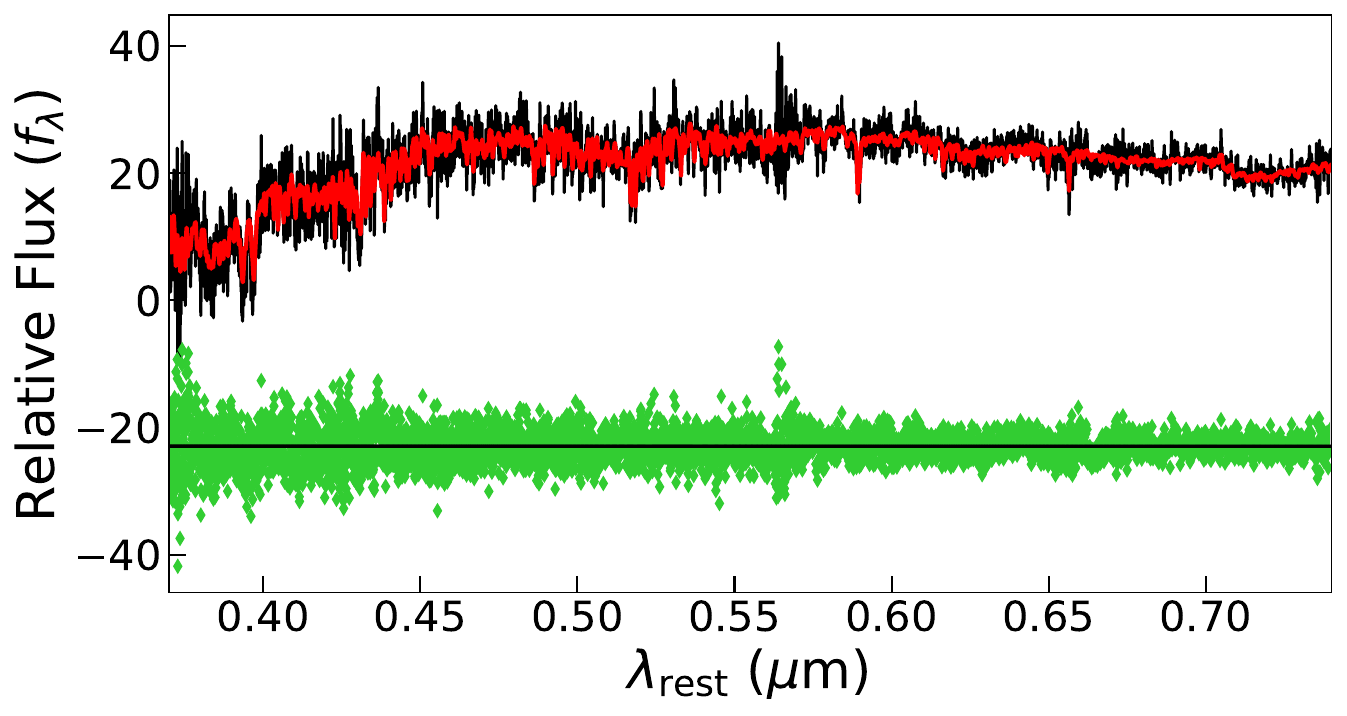}
    \caption{\textit{Left}: Best-fit broadband SED model (\textit{top}) and corresponding residuals (\textit{bottom}) with CIGALE for KUG 1208. Different curves represent the contributions from various components: attenuated stellar emission (\textit{yellow}), unattenuated stellar emission (\textit{blue}), nebular emission (\textit{green}), dust emission (\textit{red}), and AGN emission (\textit{orange}). The \textit{black} curve is the total model spectrum. The modeled and observed fluxes are overlaid as \textit{red} and \textit{violet} points, respectively. The reduced $\chi^2$, stellar mass of the host galaxy, and the SFR are marked.
    \textit{Right}: pPXF fit to the host galaxy spectrum from the \textsf{PyQSOFit}-decomposed \textit{DESI} offset-AGN spectrum. The \textit{black}, \textit{red}, and \textit{green} lines represent the observed spectrum, the best-fit stellar spectrum, and residuals, respectively. 
    }
    \label{app:fig:host}
\end{figure}

\end{appendix}

\end{document}